\documentclass[prl,superscriptaddress,twocolumn]{revtex4}
\usepackage{dcolumn}
\usepackage{bm}
\usepackage{amssymb}
\usepackage{amsmath}
\usepackage{amsfonts}
\usepackage[english]{babel}
\usepackage{amsmath}
\usepackage{graphicx}
\usepackage{dcolumn}
\usepackage{bm}
\usepackage{hyperref}

\RequirePackage[cdot,textstyle,amssymb]{SIunits}%
\RequirePackage{units}

\usepackage[]{graphicx}

\usepackage{epstopdf}

\begin{document}

\title{Anisotropic Dzyaloshinskii-Moriya Interaction in ultra-thin epitaxial Au/Co/W(110)}
\author{Lorenzo Camosi} \email{lorenzo.camosi@neel.cnrs.fr}
\affiliation{Univ.~Grenoble Alpes, CNRS, Institut N\'eel, F-38000 Grenoble, France}
\author{Stanislas Rohart}
\affiliation{Laboratoire de Physique des Solides, Universit\'{e} Paris-Sud, CNRS UMR 8502, F-91405 Orsay Cedex, France}
\author{Olivier Fruchart}
\affiliation{Univ.~Grenoble Alpes, CNRS, CEA, Grenoble INP, SPINTEC, F-38000 Grenoble, France}
\affiliation{Univ.~Grenoble Alpes, CNRS, Institut N\'eel, F-38000 Grenoble, France}
\author{Stefania Pizzini}
\affiliation{Univ.~Grenoble Alpes, CNRS, Institut N\'eel, F-38000 Grenoble, France}
\author{Mohamed Belmeguenai}
\author{Yves Roussign\'e}
\affiliation{LSPM-CNRS, Universit\'{e} Paris XIII-Sorbonne Paris Cit\'{e}, F-93430 Villetaneuse, France}
\author{Andre\"{i} Stashkevich}
\affiliation{LSPM-CNRS, Universit\'{e} Paris XIII-Sorbonne Paris Cit\'{e}, F-93430 Villetaneuse, France}
\affiliation{International laboratory ``MultiferrLab'', ITMO University, St. Petersburg, Russia}
\author{Salim Mourad Cherif}
\affiliation{LSPM-CNRS, Universit\'{e} Paris XIII-Sorbonne Paris Cit\'{e}, F-93430 Villetaneuse, France}
\author{Laurent Ranno}
\author{Maurizio de Santis}
\author{Jan Vogel}
\affiliation{Univ.~Grenoble Alpes, CNRS, Institut N\'eel, F-38000 Grenoble, France}

\begin{abstract}
We have used Brillouin Light Scattering spectroscopy to independently determine the in-plane Magneto-Crystalline Anisotropy and the Dzyaloshinskii-Moriya Interaction (DMI) in out-of-plane magnetized Au/Co/W(110). We found that the DMI strength is 2-3 times larger along the \textit{bcc}$[001]$ than along the \textit{bcc}$[\overline{1}10]$ direction. We use analytical considerations to illustrate the relationship between the crystal symmetry of the stack and the anisotropy of microscopic DMI. Such an anisotropic DMI is the first step to realize isolated elliptical skyrmions or anti-skyrmions in thin film systems with $C_{2v}$ symmetry.
\end{abstract}

\maketitle

\section{Introduction}
\begin{figure}[h]
  \begin{center}
    \includegraphics[width=9cm]{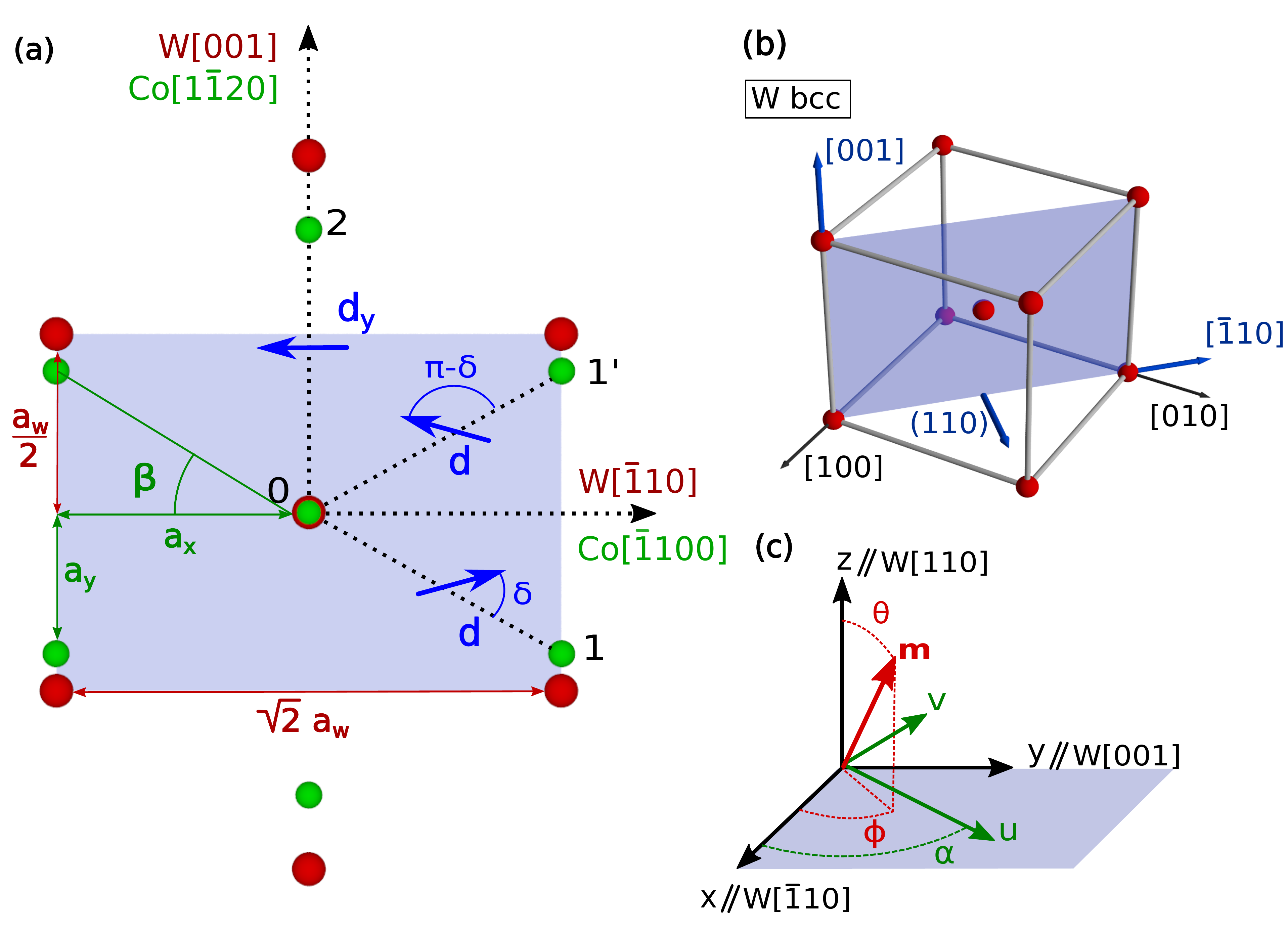}
  \end{center}
  \caption{\label{fig:epitax} \textbf{(a)} Superposition of the W(110) and the strained Co (0001) surfaces with the Nishiyama-Wassermann relationship  \textbf{(b)} Tungsten bcc unit cell with the (110) surface highlighted  \textbf{(c)} Illustration of the geometry and notation used to describe the magnetization ( $\theta; \phi$) and the directions ($ \alpha$) in the \textit{bcc}(110) crystal framework  }
\end{figure}

An anti-symmetric exchange interaction, the Dzyaloshinskii-Moriya Interaction (DMI), was theoretically predicted by Dzyaloshinskii \cite{Dzyaloshinskii1957} using symmetry arguments in bulk magnetic systems. Then Moriya \cite{Moriya1960} demonstrated the anti-symmetric spin coupling in systems with a lack of inversion symmetry, by including spin-orbit coupling in the super-exchange interaction. Fert and Levy \cite{FertLevy1980} pointed out that high spin-orbit scattering centers can break the indirect exchange symmetry. DMI presents a particular interest since it can stabilize chiral magnetic textures like skyrmions and anti-skyrmions \citep{Bogdanov2001}, magnetic solitons with a chiral vortex-like spin configuration which are characterized by a topological charge $N_{\mathrm{sk}}$.
In a continuous-field approximation $N_{\mathrm{sk}}$ can be formulated as the integral on the space $(r ,\alpha)$ that counts how many times the magnetization $\mathbf{m}(\phi(\alpha),\theta (r))$ [Fig.~\ref{fig:epitax}(c)] wraps the unit sphere \cite{Nagaosa2013}.

\small
\begin{equation}
\label{eq:topo}
N_{\mathrm{sk}}= \frac{1}{4\pi} \int \frac{d \theta}{d r} \frac{d \phi}{d \alpha} \sin \theta dr d\alpha  = W \cdot p = \pm 1
\end{equation}
\normalsize

where \textit{p} describes the direction of the core of the spin texture [$p = 1~(-1)$ if $ \theta(r~=~0) = 0~(\pi)$] and $ W =[ \phi(\alpha) ]^{\alpha=2 \pi}_{\alpha= 0}/2 \pi= \pm 1 $ is the winding number. Considering the same magnetization background, i.e. the same \textit{p} value, skyrmions ($\phi(\alpha) \propto \alpha $) and anti-skyrmions $(\phi(\alpha) \propto - \alpha )$ have opposite winding numbers and hence opposite topological charges. The spin modulation $\phi(\alpha)$, and hence the winding number, depends on the DMI symmetry that in a monocrystalline system directly arises from the crystal symmetry \cite{Bogdanov1989,kataoka1981helical}.\\

Circular skyrmions in an isotropic DMI environment have experimentally been observed in bulk systems with B20 symmetry \cite{Muhlbauer2009} and as metastable objects in ultra-thin magnetic films \cite{Romming2013,Moreau2016,Boulle2016}. Skyrmions can also display a non-cylindrical symmetry in anisotropic environments. The effect of spatially modulated exchange energy and magneto-crystalline anisotropy on the skyrmion shape has been theoretically analyzed \cite{Hagermeister2016,Gungordu2016} and experimentally investigated \cite{Hsu2016} in ultra-thin films, while a distorted skyrmion lattice \cite{Shibata2015} due to an anisotropic DMI has been evidenced in a mechanically-strained single-crystal.\\

Anti-skyrmions have been theoretically predicted in bulk systems where the $D_{2d}$ and $S_4$ \cite{Bogdanov1989} symmetry induces an anisotropic DMI with inversion of chirality between perpendicular directions. They have been theoretically investigated as metastable states at an energy higher than the skyrmion in ultrathin films with isotropic chirality \cite{Dupe2016} and in systems without DMI \cite{Koshibae2015}.\\

This paper consists of two parts. In the first part, we experimentally study thin epitaxial Co films on W(110). We use Brillouin Light Scattering (BLS) spectroscopy to show that the $C_{2v}$ crystal symmetry leads to a strong anisotropy of the DMI, with a value which is 2-3 times higher along the \textit{bcc}$[\overline{1}10]$ than along the \textit{bcc}$[001]$ direction. In the second part, we first show the relationship between the atomic DMI at the W$\backslash$Co interface and the micromagnetic DMI in a C$_{2v}$ symmetry system. Then, we analyze the spin waves and spin configurations stabilized by the anisotropic DMI energy in a general C$_{2v}$ symmetry in order to explain our BLS measurements. Finally, we show that a DMI with opposite sign along two perpendicular in-plane directions should lead to the stabilization of anti-skyrmions.

\section{Sample growth}
The sample stack is grown by pulsed laser deposition, and crystallographic properties are investigated in-situ. The ($11\overline{2}0$) surface of a commercial Al$_2$O$_3$ single crystal is used as the substrate for growing at room temperature a thin film of Mo (\unit[0.8]{\nano\meter}) followed by the deposition of a \unit[8]{\nano\meter}~thick W film. The stack is then annealed at \unit[1200]{\kelvin} for \unit[1]{h}. During this annealing the Mo underlayer promotes the selection of a unique epitaxial relationship, avoiding twins and yielding a single-crystalline film \cite{Fruchart1998}. Reflection High-Energy Electron Diffraction (RHEED), shown in the Supplemental Material \cite{supplementary}, confirms the disappearance of the W twins and the correct epitaxial relationship (Fig.~\ref{fig:epitax}). A Co film with a thickness t = 0.65\,nm is then deposited. The best condition for layer-by-layer growth was obtained by progressively warming the sample from room temperature to \unit[350]{\kelvin} while the Co thickness increases from 0 to \unit[0.65]{\nano\meter}. The immiscibility between Co and W guarantees a flat and sharp interface. RHEED and Grazing incidence X-ray diffraction (GIXRD) patterns \cite{supplementary} demonstrate the retained single crystal feature through the Nishiyama-Wassermann epitaxial relationship. The lattice misfits along the main in-plane crystallographic directions are  $\Delta a_{\textit{bcc}[\overline{1}10]}  =\frac{\sqrt{2} a_{\mathrm{W}} - \sqrt{3} a_{\mathrm{Co}}}{\sqrt{2} a_{\mathrm{W}}} = 2.98 \%  $ and  $ \Delta a_{\textit{bcc}[001]}  =\frac{a_{\mathrm{W}} - a_{\mathrm{Co}}}{a_{\mathrm{W}}} = 20.79 \%  $ where $a_{\mathrm{W}}$ and $a_{\mathrm{Co}}$ are respectively the bulk \textit{bcc} and \textit{hcp} lattice parameters. Along the $\textit{bcc}[\overline{1}10]$ direction the Co is expected to grow pseudomorphically ($a_x = \sqrt{2} /2 a_{\mathrm{W}} $), up to 10 Co monolayers (1ML $\simeq $ 0.2\,nm) \cite{Fritzsche1995}. Along the \textit{bcc}[001] direction, the misfit instead is large implying that the Co structure relaxes for a thickness between 2 and 4 ML ($a_y= 3.56/4.56  \frac{a_\mathrm{W}}{2}  $\cite{Fritzsche1995}), with $a_x$ and $a_y$ defined in Fig.~\ref{fig:epitax}. Along the \textit{bcc}[001] direction, the Co-W crystal forms a superstructure with a period of $14 a_y$ (1.5\,nm), reasonably smaller than the characteristic magnetic length scales even in ultrathin Co films. From the micromagnetic point of view the system can thus be considered uniform with averaged quantities and with a $ C_{2v}$ symmetry.

Finally, a \unit[2]{\nano\meter}-thick \textit{fcc} Au(111) cap layer is deposited in order to promote out-of-plane anisotropy and protect the stack from oxidation. This layer has a C$_{6v}$ symmetry due to the fcc Au(111) surface twins. GIXRD measurements show that the W$\backslash$Co interface is hardly modified by the capping layer \cite{supplementary} and the stressed Co layer does not significantly change its crystal symmetry. Hence we expect the contribution of the Au/Co interface to the in-plane anisotropic properties to be negligibly small.

\section{Brillouin Light Scattering spectroscopy}

Brillouin Light Scattering spectroscopy was performed in the Damon-Eshbach (DE) configuration \cite{Damon1961}. This technique is particularly suited for the study of anisotropic systems because it allows to extract the magnetic properties independently along any direction. An external magnetic field $H_{\mathrm{ext}}$ saturates the magnetization along an in-plane direction. A laser beam ($ \lambda= 532$ nm) strikes the sample in the plane perpendicular to the magnetic field with an incidence angle $0^{\circ}<\theta_{\mathrm{inc}}<60^{\circ}$ in order to vary the spin wave (SW) wave vector involved in the scattering process $k_{\mathrm{SW}} = 4 \pi \sin( \theta)/ \lambda $. We call $\alpha$ the angle between $\mathbf{k}_{\mathrm{SW}}$ (the direction along which the magnetization varies) and the \textit{bcc} $[\overline{1}10]$ crystallographic direction (Fig.~\ref{fig:epitax}). A 2x3 pass Fabry-Perot interferometer allows to analyze the back-scattered light and to study the Stokes (S) and anti-Stokes (AS) spectrum generated by the scattering process between the laser photons and the SWs for different $\alpha$ values. The BLS spectrum in systems with DMI can be separated in a symmetric $f_0 =(\vert f_{\mathrm{S}} \vert + \vert f_{\mathrm{AS}} \vert )/2  $ and an antisymmetric component $f_{\mathrm{anti}}=(\vert f_{\mathrm{S}} \vert - \vert f_{\mathrm{AS}})\vert /2 $. The study of $f_0$ with $H_{\mathrm{ext}}$ along the main crystallographic directions allows to estimate the magneto-crystalline anisotropy (MCA) constants\,$K_i$ in the direction of the applied field, while $f_{\mathrm{anti}}$ allows to extract the sign and strength of the DMI acting on a N\'eel spin cycloid along the SW wavevector.

\begin{figure}[h]
 \begin{center}
 \includegraphics[width=8.5cm]{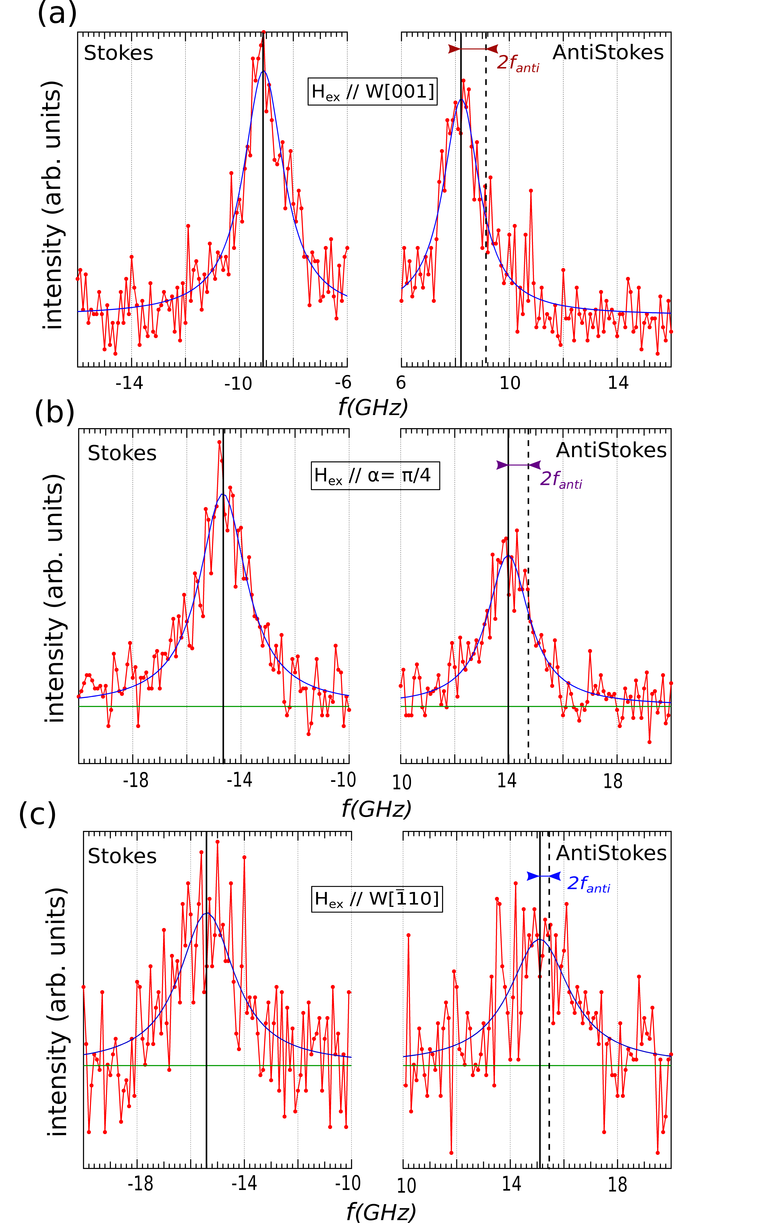}
   \end{center}
  \caption{BLS spectra on Au/Co(0.65~nm)/W(110) with $\mathbf{k}_{\mathrm{SW}}$ along the two in-plane symmetry axes. Red: experimental data. Blue line: data fit with Lorentzian functions. Green line: background fit. In the AS spectra, the distance between the continuous and dashed black lines shows the frequency shift between S and AS peaks. \textbf{(a)} BLS spectrum with $\mu_0 H_{ext}=\unit[0.6]{\tesla} $ parallel to the \textit{bcc}$[001]$ axis and $k_{\mathrm{SW}}= \unit[18.09]{\per\micro\meter}$ parallel to the \textit{bcc}$[\overline{1}10]$ axis \textbf{(b)} BLS spectrum
with $\mu_0 H_{ext}=\unit[0.6]{\tesla}$ along the direction with an angle of $\pi/4$ with respect to the bcc$[1\overline{1}0]$ axis and $k_{\mathrm{SW}}= \unit[18.09]{\per\micro\meter}$ \textbf{(c)} BLS spectrum with $\mu_0 H_{ext}=\unit[0.5]{\tesla}$ parallel to the \textit{bcc}$[\overline{1}10]$ axis and $k_{\mathrm{SW}}= \unit[18.09]{\per\micro\meter}$ parallel to the \textit{bcc}$[001]$ axis. }
\label{fig:BLSspec}
\end{figure}

The $C_{2v}$ supercrystal symmetry induces a biaxial MCA energy density that can be formulated in the second order approximation including the out-of-plane shape anisotropy ($K_{\mathrm{d}} = \frac{1}{2} \mu_0 M_\mathrm{s}^2$) :

\small
\begin{equation}
\label{eq:ani}
E_{\mathrm{anisotropy}}= -(K_{\mathrm{out}}- K_{\mathrm{d}} )\cos^2 \theta - K_{\mathrm{in}} \sin^2\theta \; \cos^2\phi
\end{equation}
\normalsize

where $\theta$ and $\phi$ describe the magnetization direction (Fig.~\ref{fig:epitax}) and $K_{\mathrm{out}}$ and $K_{\mathrm{in}}$ are the out-of-plane and the in-plane easy axis MCA constants.
The symmetric frequencies $f_0^{[001]}$ and $f_0^{[\overline{1}10]}$, when $H_{\mathrm{ext}}$ is respectively applied along $[001]$ and $[\overline{1}10]$, can be calculated \citep{Baselgia1988} as,

\begin{widetext}
\small
\begin{align}
f_0^{[001]}=& \frac{\gamma \mu_0}{2 \pi} \sqrt{[H_{\mathrm{ext}}^{[001]}-H_{\mathrm{in}}+ Jk_{\mathrm{SW}}^2+P(k_{\mathrm{SW}}t)M_\mathrm{s} ][H_{\mathrm{ext}}^{[001]}-H_{\mathrm{out}}+J k_{\mathrm{SW}}^2 -P(k_{\mathrm{SW}}t)M_\mathrm{s}]} \\
f_0^{[\overline{1}10]}=& \frac{\gamma \mu_0}{2 \pi} \sqrt{[H_{\mathrm{ext}}^{[\overline{1}10]}+H_{\mathrm{in}} + Jk_{\mathrm{SW}}^2+P(k_{\mathrm{SW}}t)M_\mathrm{s} ][H_{\mathrm{ext}}^{[\overline{1}10]}-H_{\mathrm{out}}+H_{\mathrm{in}}+J k_{\mathrm{SW}}^2 -P(k_{\mathrm{SW}}t)M_\mathrm{s}]}
\end{align}
\label{eq:f0}
\normalsize
\end{widetext}

where $\gamma$ is the gyromagnetic ratio, $J= \frac{2A}{\mu_0 M_\mathrm{s}}$ is the SW stiffness with $A$ the exchange stiffness and $M_\mathrm{s}$ the spontaneous magnetization, $P(k_{\mathrm{SW}}t)= 1 -\frac{1-\exp(-\mid k_{\mathrm{SW}}\mid t)}{\mid k_{\mathrm{SW}}\mid t}$ is a geometric factor associated to the SW dynamics with $t$ the sample thickness.
Following Eq.~(\ref{eq:ani}) we define $H_{\mathrm{out}}$ and $H_{\mathrm{in}}$ as the anisotropy fields. $H_{\mathrm{out}}$ is the magnetic field needed to saturate the magnetization along the in-plane hard axis $(\theta = \pi/2 ; \phi =\pi/2)$. $H_{\mathrm{in}}$ is the difference between the fields needed to saturate the magnetization along the in-plane easy axis $(\theta = \pi/2 ; \phi = 0)$ and the in-plane hard axis. Analyzing the spectra in Fig.~\ref{fig:BLSspec} can give a numerical estimation of the MCA constants.

\begin{figure}[h]
 \begin{center}
 \includegraphics[width=8.5cm]{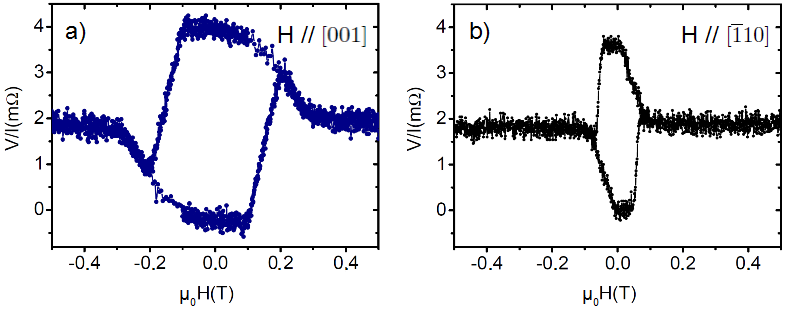}
   \end{center}
  \caption{Anomalous Hall Effect measurements of the Au/Co(0.65~nm)/W(110) sample with the magnetic field applied along the \textit{bcc}$[001]$ (a) and $\textit{bcc}[\overline{1}10]$ (b) in-plane directions.}
\label{Hall}
\end{figure}

In this work, the S-AS peaks occur for small values of $k_{\mathrm{SW}}$, i.e. $J k_{\mathrm{SW}}^2 <<  H_{\mathrm{ext}} $, so that it is possible to neglect exchange contributions to the resonance BLS peaks. The spontaneous magnetization ($M_\mathrm{s}= 1.15 \cdot 10^6$ A/m) is inferred from the out-of-plane hysteresis loop obtained with a vibrating sample magnetometer (VSM). Evaluating $f_0^{[001]}=8.53\,$GHz and $f_0^{[\overline{1}10]}= 15.24\,$GHz with respectively $\mu_0 H_{\mathrm{ext}}^{[001]}=0.6$ T and $\mu_0 H_{\mathrm{ext}}^{[\overline{1}10]}=0.5$ T we obtain $K_{\mathrm{in}} = \frac{1}{2} \mu_0  M_{\mathrm{s}} H_{\mathrm{in}}= 136 $kJ/m$^3$ ($\mu_0 H_{\mathrm{in}}$ = 0.24~T) and $ K_{\mathrm{out}}  - K_{\mathrm{d}} = \frac{1}{2} \mu_0  M_{\mathrm{s}} H_{\mathrm{out}}= 199 $kJ/m$^3$ ($\mu_0 H_{\mathrm{out}}$ = 0.35~T). Anomalous Hall Effect measurements performed on the same sample with in-plane fields along the $\textit{bcc}[\overline{1}10]$ $(\theta = \pi/2 ; \phi = 0)$ and \textit{bcc}$[001]$ $(\theta = \pi/2 ; \phi = \pi/2)$ directions give saturation fields $\mu_0(H_{\mathrm{out}} - H_{\mathrm{in}}) \approx 0.1$~T and $\mu_0 H_{\mathrm{out}} \approx 0.3$~T, in good agreement with the anisotropy values (Fig.~\ref{Hall}). Note that published results on the same system \cite{Sellmann2001} showed a comparable out-of-plane anisotropy, but a larger in-plane anisotropy.

The difference $2 f_{\mathrm{anti}}$ arises from the different effect of DMI on SW modes with opposite $\mathbf{k_{\mathrm{SW}}}$ \cite{Zakeri2010,Stashkevich2015}. In ultra-thin films DMI is the only physical phenomenon liable to break the S-AS peak symmetry \cite{Stashkevich2015}. BLS is thus particularly suited for the investigation of anisotropic DMI, especially because the extracted data are independent from any other anisotropy present in the system such as MCA, and from the strength of $H_{\mathrm{ext}} $. The SW frequency shift in a system with interfacial DMI [$ D (t)= D_s / t $)] in the DE geometry can be formulated as \cite{Zakeri2010,Udvardi2009}:

\small
\begin{equation}
2 f_{\mathrm{anti}}=  \frac{2 \gamma}{\pi} \frac{D(t)}{M_\mathrm{s}}  k_{\mathrm{SW}} = \frac{2 \gamma}{\pi}  \frac{D_s }{M } k_{\mathrm{SW}} \,.
\end{equation}
\normalsize

\begin{figure}
  \begin{center}
    \includegraphics[width=9.1cm]{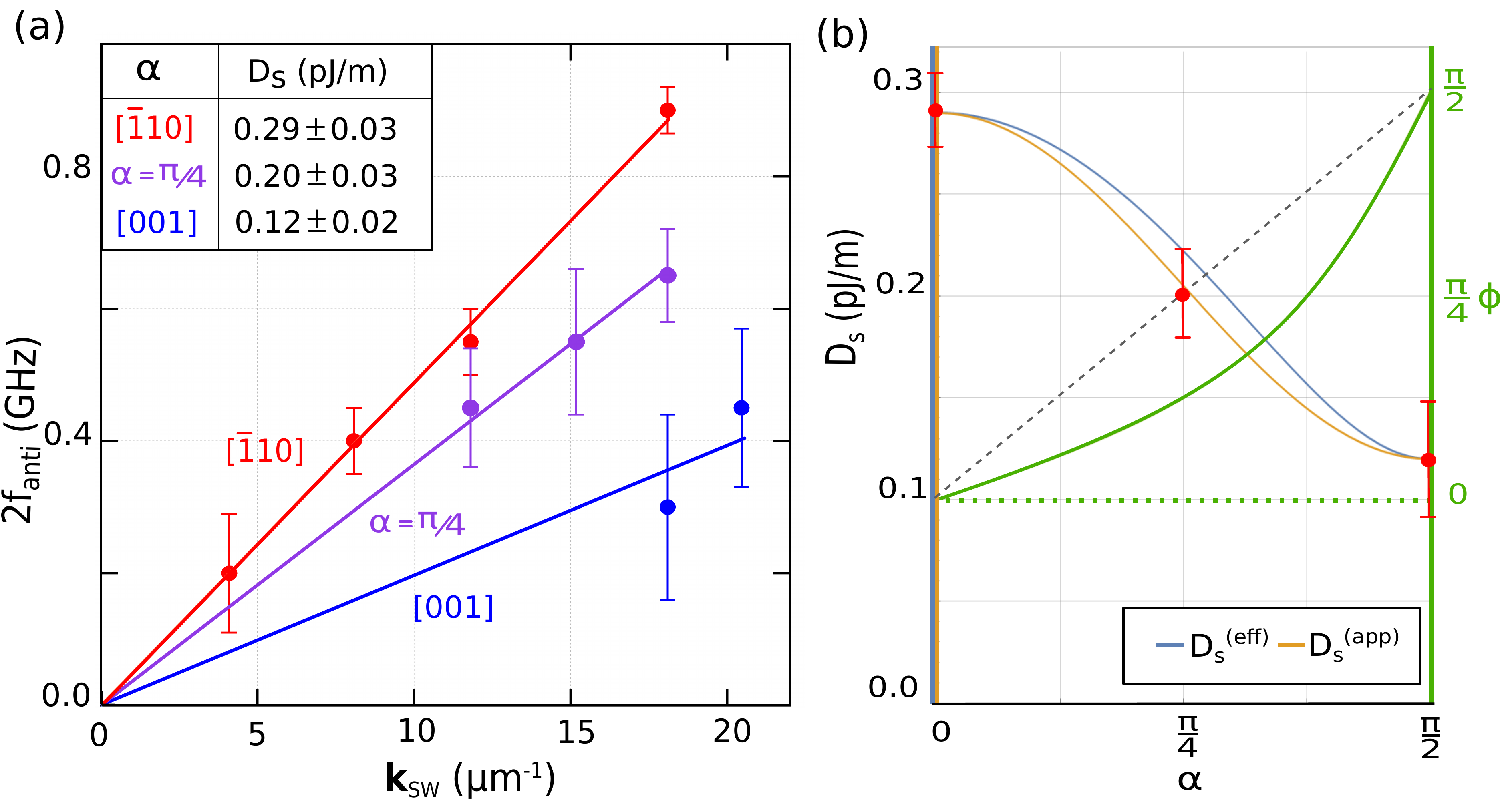}
  \end{center}
  \small
  \caption{\label{fig:DMIdf} \textbf{(a)} S-AS frequency shift ($2f_{\mathrm{anti}}$) as a function of SW wave-vector ($k_{\mathrm{SW}} $) for different in-plane directions~$\alpha$. The dots are the experimental data and the lines are linear fit yielding the DMI strength ($D_{\mathrm{s}}$). \textbf{(b)} Blue and orange lines : micromagnetic calculated $D^{(\mathrm{eff})}_{\mathrm{s}}$(Eq.~(\ref{eq:Deff})) and $D^{(\mathrm{app})}_{\mathrm{s}}$(Eq.~(\ref{eq:Dapp}))  as a function of the in-plane directions ($\alpha$); red dots: D strength evaluated from the experimental data; green line: micromagnetic calculated magnetization direction promoted by DMI (Eq.~(\ref{eq:phialp})) as a function of the crystallography directions; dashed line: N\'eel-like cycloid }
  \normalsize
\end{figure}

M, the magnetic moment per unit surface ($M = M_\mathrm{s}\,t$), is obtained directly from VSM measurements, allowing a thickness-independent determination of the DMI strength, $D_\mathrm{s}$. In Fig.~\ref{fig:DMIdf} $2 f_{\mathrm{anti}}$ is plotted as a function of $k_{\mathrm{SW}}$ along the main axes (\textit{bcc}$[001]$ ; \textit{bcc}$[\overline{1}10]$) and along an intermediate direction ($\alpha= \pi/4$). The points in the plot are extracted from the center of the Lorentzian distribution used to fit the S and AS peaks (Fig.~\ref{fig:BLSspec}). The error bars ($\delta f$) are obtained by a Levenberg-Marquardt error algorithm. The difference in the magnitude of errors (Fig.~\ref{fig:DMIdf}) between the in-plane directions is related to an instrumental issue that leads to a decrease of the signal-to-noise ratio in the BLS spectra when the magnon frequency increases (Fig.~\ref{fig:BLSspec}).

The plot in Fig.~\ref{fig:DMIdf} demonstrates that along all directions 2f$_{\mathrm{anti}}$ has a positive value, showing that the DMI promotes a clockwise spin chirality. Such a clockwise chirality (positive value of D) is in agreement with results found for sputtered MgO/CoFeB/W samples \cite{Torrejon2014} and is opposite to the chirality in AlOx/Co/Pt films \cite{Belmeguenai2015,Pizzini2014}.
Moreover, the DMI is strongly anisotropic. In the table in Fig.~\ref{fig:DMIdf}(a) the values of $D_\mathrm{s}$ along different crystallographic directions are shown. The DMI strength is a factor 2 to 3 higher along the \textit{bcc}$[\overline{1}10]$ than along the \textit{bcc}$[001]$, even taking into account the large error bar especially along the [001] direction. This difference is also confirmed by the intermediate value found for the DMI strength for SWs propagating along the intermediate angle $\alpha= \pi/4$.\\

\section{DMI and crystal symmetry : micromagnetic calculations}
Experimentally we have thus found a 2-3 times larger DMI along the $[\overline{1}10]$ than along the $[001]$ in-plane directions. In order to understand the relation between the crystal symmetry, the micromagnetic DMI anisotropy and the symmetry of the spin modulation $\phi(\alpha)$ we developed micromagnetic calculations. Our approach does not aim at the quantitative evaluation of the DMI, but allows illustrating how the C$_{2v}$ crystal symmetry sets constraints on the atomic DMI vectors $d_{ij}$ and how to obtain the anisotropic micromagnetic D constants. It is valid if the analyzed magnetic configurations have a characteristic length ($l$) much larger than the supercell parameter ($14 a_x$). Indeed it allows considering averaged $\langle \mathbf{d}_{ij} \rangle$ on all the superlattice and describing the magnetization in a continuous medium approach. The symmetry in a C$_{2v}$ crystal is not high enough to set uniquely the $\langle \mathbf{d}_{ij} \rangle$ vectors \cite{Moriya1960} but imposes their directions in the crystal plane and their mutual relationships \cite{supplementary}. The $\langle \mathbf{d}_{02} \rangle$ is perpendicular to its bond whereas $\langle \mathbf{d}_{01} \rangle$ and $\langle \mathbf{d}_{01^{\prime}} \rangle$ have the same strength $d$ and supplementary angles ($\delta_{01}+  \delta_{01^{\prime} }  = \pi  $) with respect to their bond (see Fig.~\ref{fig:epitax}). Using the notation of the Lifshitz invariants $L^{(i)}_{jk}= m_j \frac{\partial m_k}{\partial i}-m_k \frac{\partial m_j}{\partial i} $ the micromagnetic DM energy can be written:

\small
\begin{equation}
\label{eq:EDMLcon}
E_{DM} =  -  \int \left(  D^{(x)}_{\mathrm{s}} L_{xz}^{(x)}+ D^{(y)}_{\mathrm{s}} L_{yz}^{(y)}\right) d^2r
\end{equation}
\normalsize

with $ D^{(x)}_{\mathrm{s}} = \frac{d}{a} \frac{\sin( \beta + \delta_{01})}{\sin \beta}$ and $ D^{(y)}_{\mathrm{s}} = \frac{2d}{a} \left[ \frac{\cos( \beta + \delta_{01})}{\cos \beta}  -\frac{\langle d_{02} \rangle}{d} \frac{1}{\cos \beta} \right] $ (see Fig.~\ref{fig:epitax}). These relations show that knowing the crystal structure and the micromagnetic DMI is not sufficient to determine all $\langle \mathbf{d}_{ij} \rangle$ vectors.\\

In order to understand the $\phi(\alpha)$ allowed in a general C$_{2v}$ system we formulate the DMI energy of a uni-dimensional spin modulation propagating along $\widehat{u}$ in a basis $(\widehat{u} ,\widehat{v},\widehat{z})$, turned at an angle $\alpha = (\widehat{x},\widehat{u})$ with respect to the crystal basis.

\small
\begin{align}
\label{eq:EDMuv}
E_{DM}(\alpha) =  -  \int &  \left[ \cos^{2}(\alpha) D^{(x)}_{\mathrm{s}} + \sin^{2}(\alpha) D^{(y)}_{\mathrm{s}} \right]  L_{uz}^{(u)}  d^2r  \notag \\
 -  \int  & \left(  D^{(x)}_{\mathrm{s}}- D^{(y)}_{\mathrm{s}}  \right) \cos(\alpha)\sin(\alpha) L_{vz}^{(u)}  d^2r.
\end{align}
\normalsize

$E_{DM}(\alpha)$ presents two different types of Lifshitz invariants that describe a DMI stabilizing different spin modulations \cite{Bogdanov2001}. The first term  $L_{uz}^{(u)}$ describes the well known result of an interfacial DMI promoting a N\'{e}el cycloid. The second term $L_{vz}^{(u)}$ evidences that the interfacial DMI can stabilize a Bloch helicoid. This component vanishes along the main axes and has maxima proportional to the difference of the DMI constants $ ( D^{(x)}_{\mathrm{s}}- D^{(y)}_{\mathrm{s}})$ when $\alpha= \pi/4+ n\pi/2$. It means that in a general C$_{2v}$ system the DMI promotes N\'eel cycloids along the main axes and a mixed configuration between a N\'eel cycloid and a Bloch helicoid along the intermediate directions.\\

Eq.~(\ref{eq:EDMuv}) allows us first to calculate the apparent DMI constant $D^{(app)}_{\mathrm{s}}$ \cite{supplementary}, defined as the DMI component acting on the DE spin wave, as a function of the in-plane propagation direction. In the DE geometry a SW propagating along $\mathbf{\hat{u}}$ can be described as $\mathbf{m}(u) = \mathbf{M}+\delta\mathbf{m}(u,t)$, with $\mathbf{M} \parallel \widehat{v}$ is imposed by $\mathbf{H}_{\mathrm{ext}}$. The component $\delta \mathbf{m}(u,t)$, which represents the magnetization varying part is a N\'eel cycloid lying in the $(\widehat{u},\widehat{z})$ plane. Then $D^{(app)}_{\mathrm{s}}$ calculated from the DMI energy density of the SW reads, as a function of $\alpha$~\cite{supplementary}:

\small
\begin{equation}
 D^{(app)}_{\mathrm{s}}= D^{(x)}_{\mathrm{s}} \cos^2 \alpha + D^{(y)}_{\mathrm{s}} \sin^2 \alpha   \label{eq:Dapp}
\end{equation}
\normalsize

Eq.~(\ref{eq:Dapp}), plotted in Fig.~\ref{fig:DMIdf}(b), matches well the experimental data. It corresponds to the first part only of eq.~(\ref{eq:EDMuv}), due to the fact that $\delta \mathbf{m}$ describes a cycloid. Measuring the second part of eq.~(\ref{eq:EDMuv}) would require to change the measurement geometry and turn the optical plane by $\pi/2$ to get SWs propagating along the field direction ($\mathbf{M}$ along $\widehat{u}$ and $\delta \mathbf{m}(u,t)$ in the $(\widehat{v},\widehat{z})$ plane) with $\delta \mathbf{m}$ describing a helicoid.

\section{Skyrmions and Anti-skyrmions}

\begin{figure}
 \begin{center}
    \includegraphics[width=9.1cm]{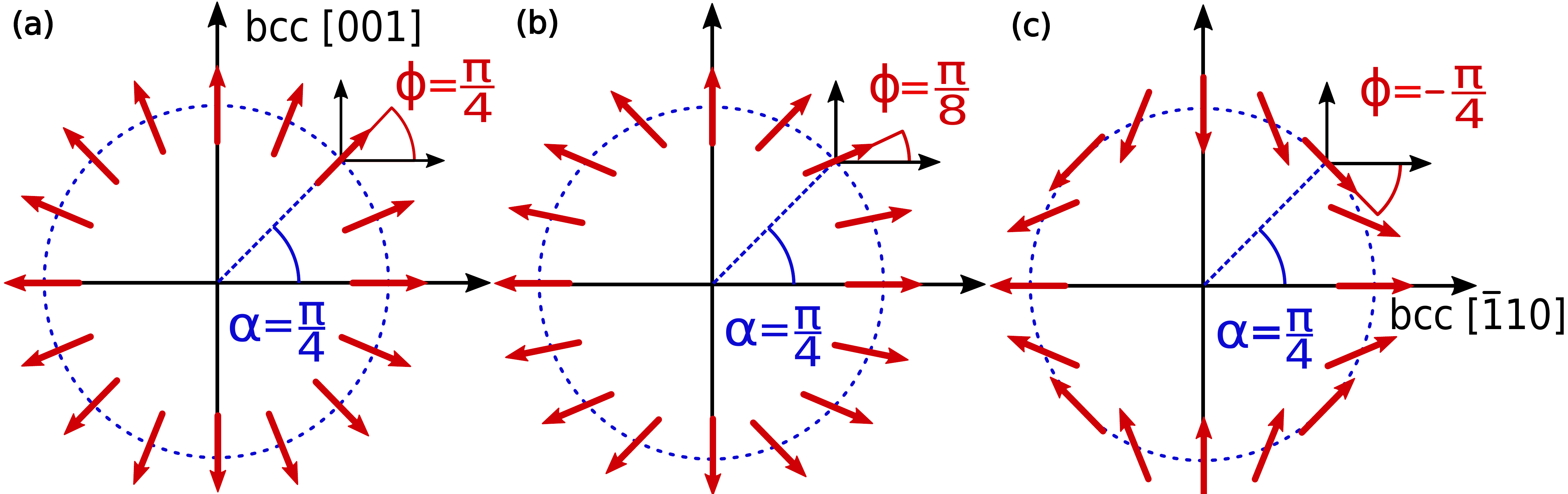}
  \small
  \caption{\label{fig:polarplot} Polar plot of magnetization direction ($\phi$) promoted by DMI as a function of the in plane direction of variation $\alpha$ (Eq.~(\ref{eq:phialp})) for different $( D^{(x)}_{\mathrm{s}};  D^{(y)}_{\mathrm{s}})$ values: (a) $ D^{(x)}_{\mathrm{s}} =  D^{(y)}_{\mathrm{s}}$ (b) $ D^{(x)}_{\mathrm{s}} = 2.5 D^{(y)}_{\mathrm{s}} $ (c) $ D^{(x)}_{\mathrm{s}} = -  D^{(y)}_{\mathrm{s}}$ }
  \normalsize
  \end{center}
\end{figure}

The competition between the first and second parts in eq.~(\ref{eq:EDMuv}) implies that along an arbitrary direction, spin spirals (or equivalently domain walls) may be intermediate between N\'eel and Bloch spirals. Writing $\phi$ as the angle between the spiral modulation plane and the $\widehat{x}$ axis, we minimize the DMI energy to find the optimum modulation plane. As a function of the propagation direction, we obtain
\small
\begin{align}
 \tan\phi =& \left( \frac{ D^{(y)}_{\mathrm{s}}}{ D^{(x)}_{\mathrm{s}}} \right) \tan\alpha \label{eq:phialp}
 \end{align}
\normalsize
with an effective DMI constant that maximizes the DMI energy gain.
\small
\begin{align}
D_s^{eff}=   D^{(x)}_{\mathrm{s}} \cos\alpha & \, \cos\left[ \arctan\left( \frac{ D^{(y)}_{\mathrm{s}}}{ D^{(x)}_{\mathrm{s}}} \tan \alpha \right)\right] \nonumber  \\
+  D^{(y)}_{\mathrm{s}} \sin\alpha & \,\sin\left[ \arctan \left( \frac{ D^{(y)}_{\mathrm{s}}}{ D^{(x)}_{\mathrm{s}}}\tan \alpha \right )\right] \label{eq:Deff}
 \end{align}
\normalsize
Hence setting $ D^{(x)}_{\mathrm{s}}=2.5 D^{(y)}_{\mathrm{s}} $ it is possible to obtain the $D^{(\mathrm{eff})}_{\mathrm{s}}(\alpha)$ for Au/Co/W(110) (Figs. \ref{fig:BLSspec}(b) and \ref{fig:polarplot}(b)). As predicted from Eq.~(\ref{eq:EDMuv}), along the main axes $D^{(\mathrm{eff})}_{\mathrm{s}} = D^{(\mathrm{app})}_{\mathrm{s}}$ and the spin spiral is purely N\'eel ; the largest mismatch between $D^{(\mathrm{eff})}_{\mathrm{s}}$ and $D^{(\mathrm{app})}_{\mathrm{s}}$ occurs along $\alpha = \pi/4$.

The discussion can be generalized for different $D^{(x)}_\mathrm{s}/D^{(y)}_\mathrm{s}$ ratio and we emphasize two interesting cases. First, setting isotropic conditions ($D^{(x)}_\mathrm{s} = D^{(y)}_\mathrm{s}$) we obtain the well-known result of a DMI stabilizing only N\'eel spirals ($\phi(\alpha) = \alpha$). On the other extreme, $D^{(x)}_\mathrm{s} = - D^{(y)}_\mathrm{s}$ implies $\phi(\alpha) = -\alpha$ and so, N\'eel cycloids are stabilized along the main crystallographic directions and purely Bloch helicoids are stabilized at $\alpha= \pi/4+ n\pi/2$. Considering localized textures such as bubbles, the energy minimization remains valid and so, different type of textures can be expected, as depicted in Fig.~\ref{fig:polarplot} for respectively $D^{(x)}_\mathrm{s}/D^{(y)}_\mathrm{s} = 1$, 2.5 and -1 \cite{Zimmermann2014}. Considering the winding number $W =[ \phi(\alpha) ]^{\alpha=2 \pi}_{\alpha= 0}/2\pi$, the first two textures depict skyrmions (although in the second case we may expect distortions) with $W = 1$, while the third case, with $W = -1$, has the signature of an anti-skyrmion \cite{Nagaosa2013}.

In order to experimentally achieve a system with opposite signs of $D$ along two perpendicular in-plane directions, one possibility would be to replace the Au cover layer in our sample with a heavy metal (HM) layer inducing a DMI at the HM/Co interface which is opposite in sign to the DMI at the Co/W interface. This DMI could be isotropic and should have a DMI strength in between the values found along the \textit{bcc}$[\overline{1}10]$ and along the \textit{bcc}$[001]$ directions in the Au/Co/W(110) system. Another possibility would be to use a system suggested recently in a theoretical paper discussing anisotropic DMI and anti-skyrmions in Fe/W(110) \cite{Hoffmann2017}.\\

\section{Conclusions}
We have investigated DMI in an out-of-plane magnetized epitaxial Au/Co(0.65 nm)/W(110) trilayer. The DMI in this system promotes a clockwise chirality of the spin modulation with a DMI strength $2-3$ times larger along \textit{bcc}$[\overline{1}10]$ than along \textit{bcc}$[001]$. This anisotropy arises from the $C_{2v}$ symmetry of the Co/W(110) stack. We used a micromagnetic model to highlight the link between the atomic DMI at the Co/W(110) interface, based on its expected superlattice, and the resulting micromagnetic anisotropic DMI. The DMI is expected to give rise not only to N\'eel cycloids, but to mixed cycloid/helicoid textures [Fig.~\ref{fig:polarplot}(b)]. The experimental evidence of a strongly-anisotropic DMI is the first important step for the stabilization in a magnetic thin film of deformed isolated skyrmions and antiskyrmions.\\

We express our thanks to Philippe David and Val\'erie Guisset for the crucial support in the sample growth. We acknowledge a grant from the Laboratoire d'excellence LANEF in Grenoble (ANR-10-LABX-51-01), and support from the ANR (project ANR-14-CE26-0012 ULTRASKY) and from the Government of the Russian Federation (Grant 074-U01). We thank A. Wartelle, S. Bl\"ugel, B. Zimmermann and M. Hoffmann for helpful discussions and U. R\"ossler for useful correspondence. We acknowledge the European Synchrotron Radiation Facility and the french CRG-IF beamline for providing beamtime.

\newpage
\begin{widetext}
\part{Supplemental Materials}

\section{Sample growth}
 \label{sec:suppI}
The high nucleation density makes Pulsed Laser Deposition (PLD) an excellent technique for the layer-by-layer growth of epitaxial systems. In our set-up, the laser source is a frequency-doubled Nd:YAG laser ($\lambda= 532$ nm) with a pulse duration of approximately 10 ns, a 3 W maximum average power and 10 Hz frequency.

\begin{figure}[h]
  \begin{center}
    \includegraphics[width=17cm]{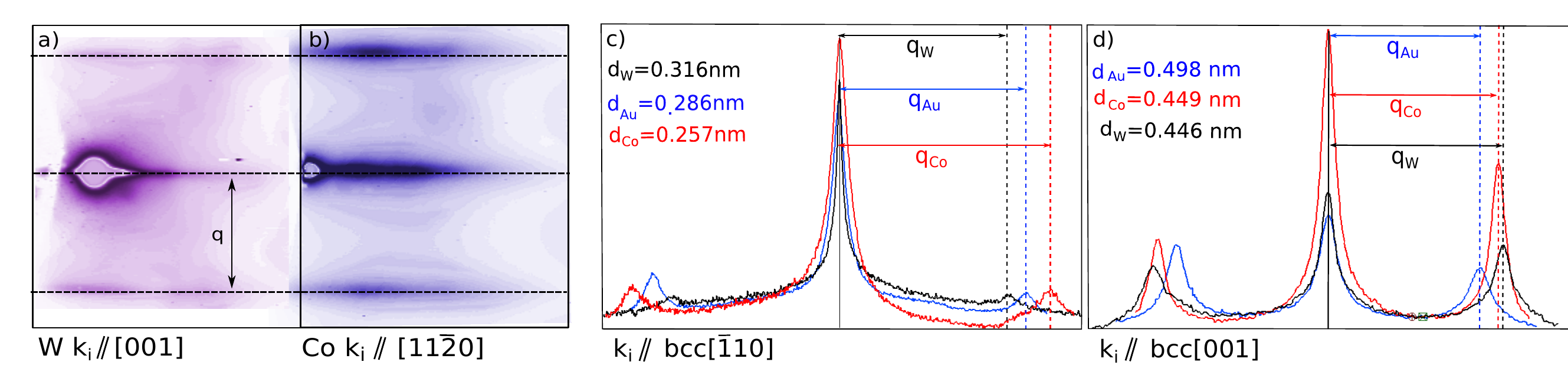}
  \end{center}
  \caption{\label{fig:WCoR} a) RHEED pattern from the W($110$) surface with the electron beam parallel to the $[001]$ direction. b) RHEED pattern from the Co surface with the electron beam parallel to the [$11\overline{2}0$] direction (parallel to the bcc $[001]$ direction). c) d) Plot of the RHEED intensity from the Gold (blue), Cobalt (red) and Tungsten (black) surfaces with the electron beam along the bcc$[\overline{1}10]$ in  c) and along the bcc[001] in d) }
\end{figure}

The Co grows on the bcc W(110) surface following the Nishiyama-Wassermann orientation, i.e. with a unique epitaxial relationship and the Co [$11\overline{2}0$] direction parallel to the W[001] and the Co $[\overline{1}100]$ parallel to the W$[\overline{1}10]$. The RHEED diffraction pattern streaks,  Fig.~\ref{fig:WCoR}(c), allow deriving the lattice parameter $a$ in the direction perpendicular to the electron beam:

\begin{equation}
a= \frac{\lambda L}{ q}
\end{equation}

where L is the distance between the detector and the sample and $\lambda$ is the wave length of the incident beam in the relativistic formulation. The value of $a^*$, reciprocal of $a$, can be used to determine the strain of the Co crystal.
Fig.~\ref{fig:WCoR}(d) shows the RHEED pattern for the Co, W and Au surfaces with the electron beam along the bcc$[001]$ direction. We can consider pseudomorphic growth of the Co along the bcc$[\overline{1}10]$ direction of the W substrate even if there is a difference between the Co and W pattern streaks. Indeed the distance between the streaks in the RHEED pattern strongly depends on the geometry of the beam reflection and the geometrical conditions between the two measurements could have slightly changed. This effect can generate an intrinsic error in the position of the pattern streaks.
It is possible to evaluate the value of the strain ($ \epsilon_{[1\overline{1}0]} = -2.86\% $ ), which is comparable with the values found for the same system in literature \cite{Sellmann2001b,Fritzsche1995b,Pratzer2003,Vandermerwe1994}.
The RHEED pattern with the electron beam along the bcc$[\overline{1}10]$ (Fig.~\ref{fig:WCoR}(c)) shows the presence of a relaxed Co structure. Indeed the large misfit between the W and the Co atomic parameters ($a_{\mathrm{ W[001]bcc}}-a_{\mathrm{ Co[001]bcc}} = 0.66 \mathrm{\AA}$ ) does not allow a pseudomorphic growth. The Co thus grows with a relaxed structure and a fixed atomic distance ratio with respect to the tungsten substrate. In literature, High Resolution Low Energy Electron Diffraction (HR-LEED) was performed on Co/W(110) for different Co thickness reporting a ratio of $\frac{a_{\mathrm{Co}}}{a_{\mathrm{W}}}  = 3.56/4.56  \frac{a_{\mathrm{W}}}{2} = 0.78$ between 2 and 4 MonoLayers (ML) \cite{Fritzsche1995b}. Our experimental data show a Co/W atomic distance ratio in agreement with this value ($\frac{a_{\mathrm{Co}}}{a_{\mathrm{W}}} = 0.81$). Hence the Co-W crystals produce a superstructure with a twofold symmetry, a period of $14 a_y$ ($14 a_y - 11 a_{\mathrm{W}} /2=2 $pm) along the W$[001]$ axis and one W atomic distance $a_x$ along the W$[1\overline{1}0]$ as shown in Fig.~\ref{fig:super}.

The system is capped with a thin film of Au. The Co(0001) symmetry allows the epitaxial growth of a fcc(111) Au crystal. It grows in its relaxed configuration due to the big mismatch of the lattice parameters.\\

Grazing incidence X-ray diffraction measurements were performed at the BM32 beamline of the European Synchrotron Radiation Facility, on the capped Au/Co/W(110) multilayer with a homogeneous Co layer of 3 ML thickness.

\begin{figure}[h!]
  \begin{center}
    \includegraphics[width=11cm]{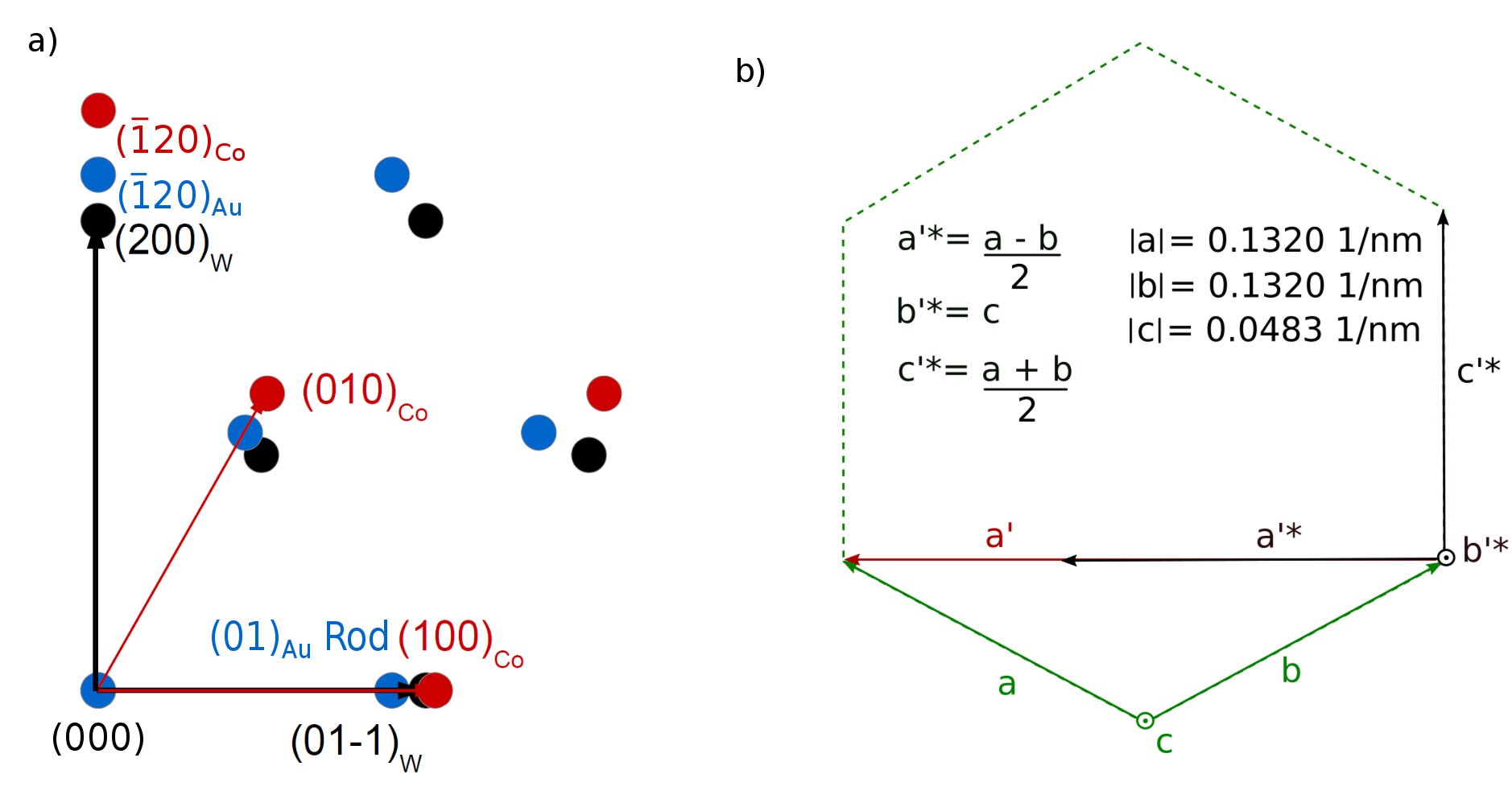}
  \end{center}
  \caption{\label{fig:Xrrel} a) Bragg peaks for \textit{fcc} Au(111)/Co(0001)/\textit{bcc} W(110) crystals in their epitaxial relationship  b) Sketch of the reciprocal framework fixed on the Al$_2$O$_3$ crystal used to describe the Bragg peaks and values of the lattice parameter for a Al$_2$O$_3$ crystal. }
\end{figure}

It is possible to describe the Bragg peaks in a reciprocal framework fixed on the Al$_2$O$_3$ crystal. The Al$_2$O$_3$ crystal has a C$_{3v}$ symmetry whereas the Au/Co/W(110) has a C$_{2v}$ symmetry. Then in order to better describe the W, Co and Au peaks we define a new framework with perpendicular axes as shown in Fig.~\ref{fig:Xrrel}.
For the W we use bcc indexes whereas for the Co and Au we use the hexagonal surface unit cell. We can formulate an expression for the points in the reciprocal space:

\begin{equation}
\mathbf{Q}=( H a'^* , K b'^* , L c'^*)  \qquad \vert \mathbf{Q}\vert = \sqrt{(H \sqrt{3} \vert a \vert)^2 +(K \vert c \vert)^2 +(L\vert a \vert)^2}
\end{equation}

with the reciprocal lattice parameter defined in Fig.~\ref{fig:Xrrel}. The momentum transfer modulus was scanned in the surface plane ($Q_z=0.08 \AA^{-1}$) along both the \textit{bcc}[001] and \textit{bcc}[$\overline{1}$10] directions. In the former case, shown in Fig.~\ref{fig:Xr001}, three Bragg peaks are observed corresponding to W(200), Au$(\overline{1} 2 0)$ and Co$(\overline{1} 2 0)$ reflections, respectively. The registry position of the cobalt layer along the \textit{bcc}$[\overline{1}10]$ direction is confirmed by the scan of Fig.~\ref{fig:Xr110}. In this case only one additional peak is observed, attributed to the relaxed Au layer. The Co(100) peak merges with the W$(\overline{1}10)$ one. Angular scans show that the main crystallographic axes of the cobalt film are aligned with the tungsten ones. Defining $\beta$ as the angle between the Co bonds 01 and 01' (Fig.~\ref{fig:super}) it is possible to determine the distortion of the Co crystal. This angle can be calculated from the position of the Co (100) and (010) peaks ($\beta= 0.51 $). We can conclude that the Co/W interface is hardly modified by the capping layer.

\begin{figure}[h!]
  \begin{center}
    \includegraphics[width=8cm]{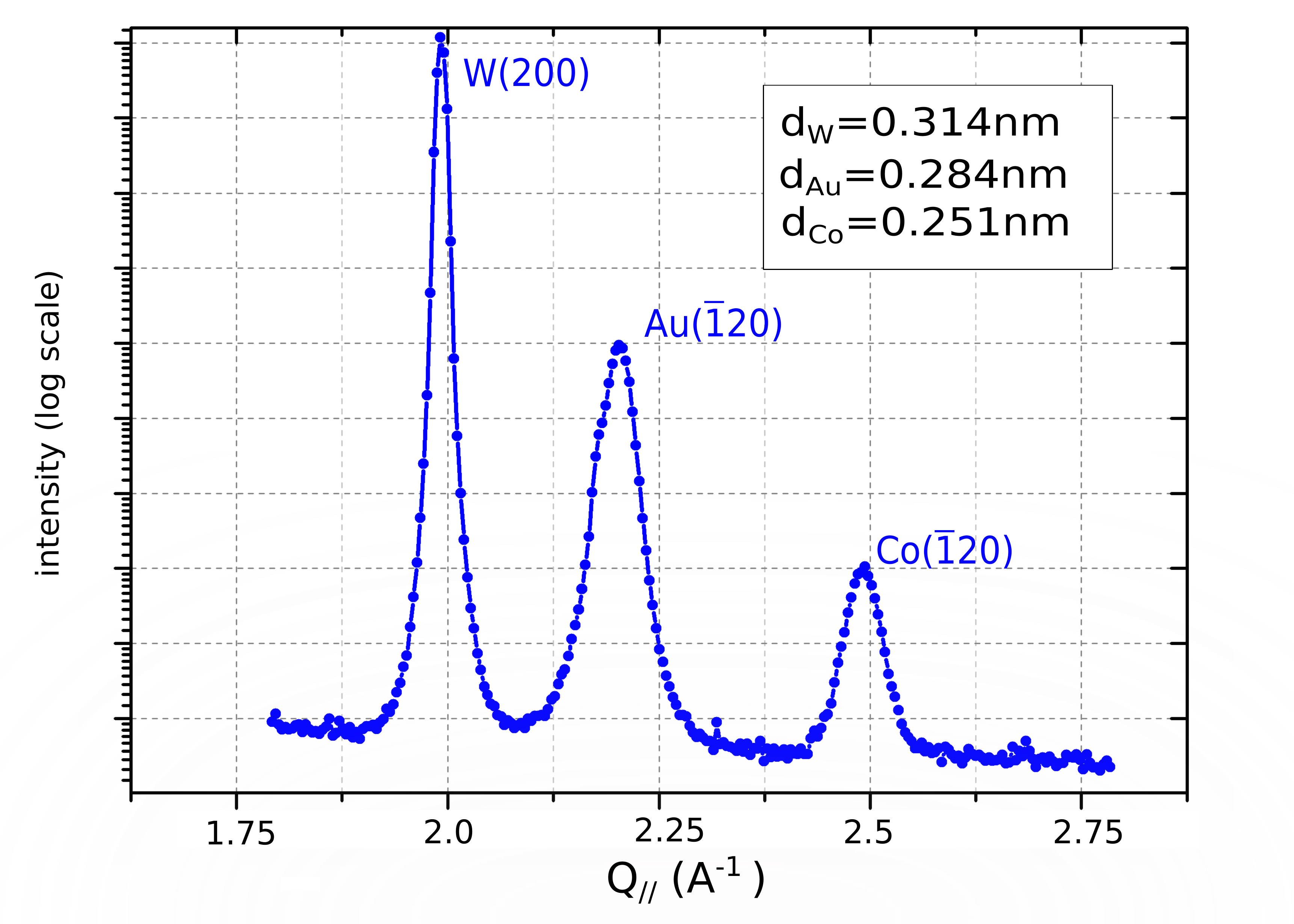}
  \end{center}
  \caption{\label{fig:Xr001}GIXRD measurements performed scanning the momentum transfer parallel to the surface plane, along the \textit{bcc}(001) direction  }
\end{figure}

\begin{figure}[h!]
  \begin{center}
    \includegraphics[width=8cm]{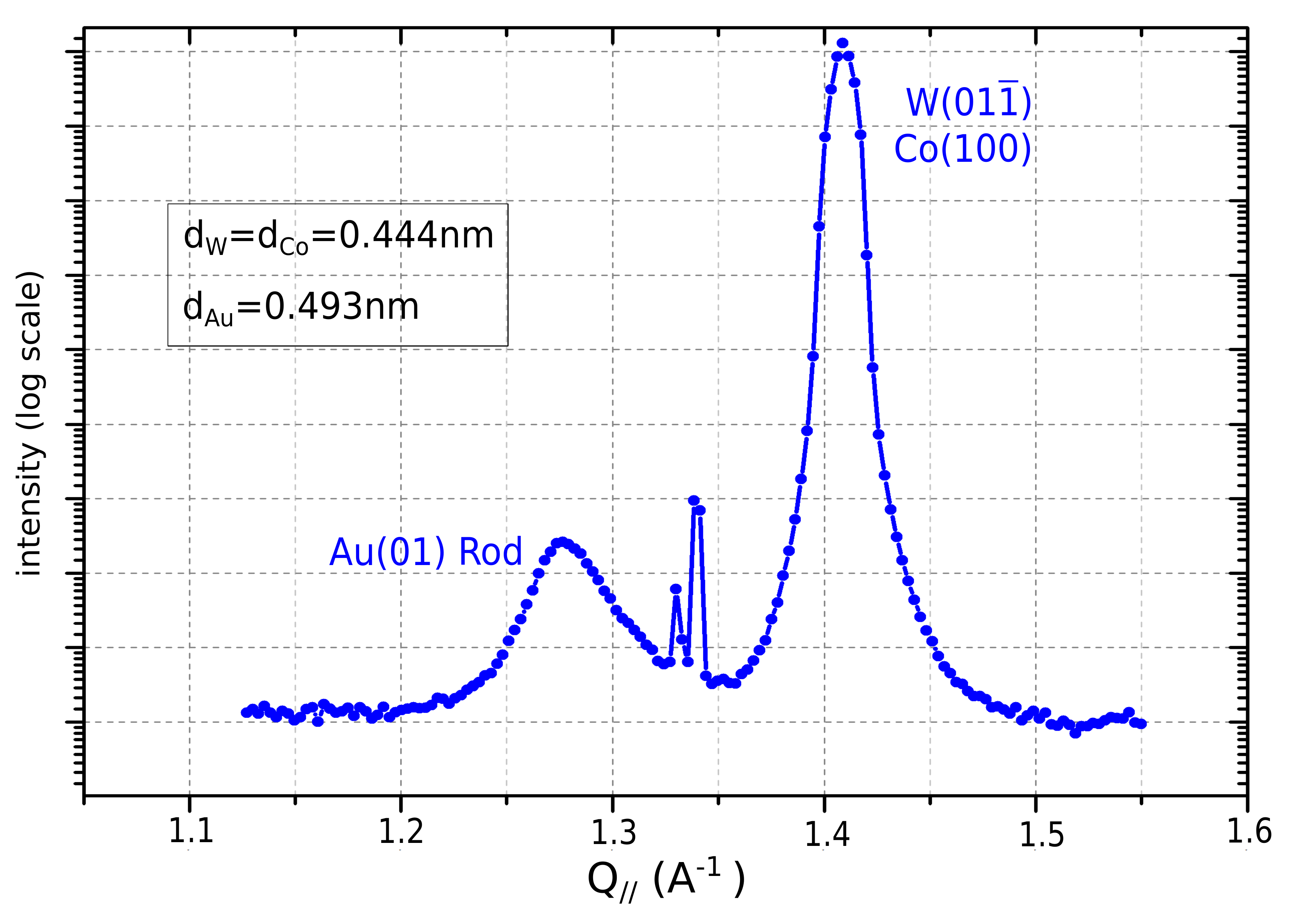}
  \end{center}
  \caption{\label{fig:Xr110} Scan parallel to the \textit{bcc}$(\overline{1}10)$ direction }
\end{figure}

We also prepared a Co/W(110) sample with a thickness gradient of the Co layer. The layer thickness is calculated a priori using an in-situ quartz crystal microbalance placed before the deposition at the sample position. Fig.~\ref{fig:STM} shows the STM pictures taken along a Co wedge. The Co islands, as shown in Fig.~\ref{fig:STM}(f), have the height of the Co interplanar distance ($2 \mathrm{\AA} $) and their lateral size increases for increasing values of the Co thickness. The growth is not perfectly layer-by-layer, since in Fig.~\ref{fig:STM}(b) it is possible to detect three atomic levels. However the sample can be considered to have a homogeneous thickness from the magnetic point of view because the characteristic exchange length ($l_{ex}$) is comparable with the average distance between the islands \cite{Andrieu2001}. These images allow us to have an extra confirmation of the sample thickness. Indeed it is possible to calculate the ratio of surface covered by islands as a function of the position in the wedge. The data, as in Fig.~\ref{fig:STM}(e), are fitted with a Gaussian function for each atomic step. The thickness in ML ($t= n + CR$) is calculated via the islands coverage ratio $CR = \frac{I(n+1)}{I(n+1)+I(n)} $, where $I(n)$ is the Gaussian integral for a given n layer. The higher step linewidth function is fixed using the value of the lower step function. This allows to avoid the apparent broadening of the island size due to the STM tip shadow effect.

\begin{figure}[h]
  \begin{center}
    \includegraphics[width=14cm]{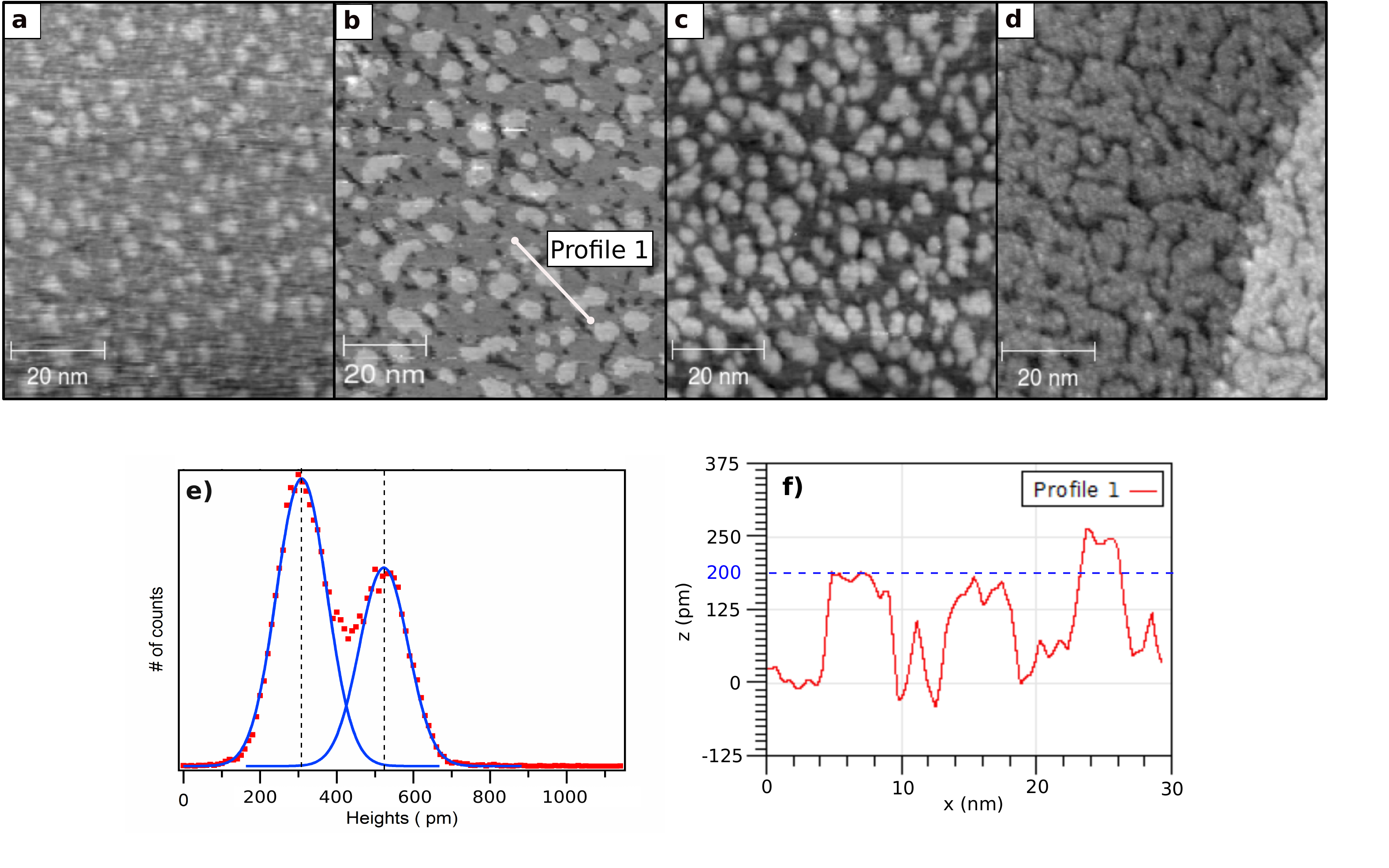}
  \end{center}
  \caption{\label{fig:STM} STM pictures of Co islands during a quasi-layer-by-layer deposition in different positions along a Co wedge. a) 1.15 ML b) 1.45 ML c) 1.62 ML d) 1.93 ML are the thickness of the Co layer that can be calculated studying the coverage ratio of Co islands. e) Plot of the heights of the islands as a function of the STM picture (b). f) plot of the islands height along the profile 1 in the STM picture (b)}
\end{figure}

\section{Out-of-plane magnetization}

\begin{figure}[h]
  \begin{center}
    \includegraphics[width=8cm]{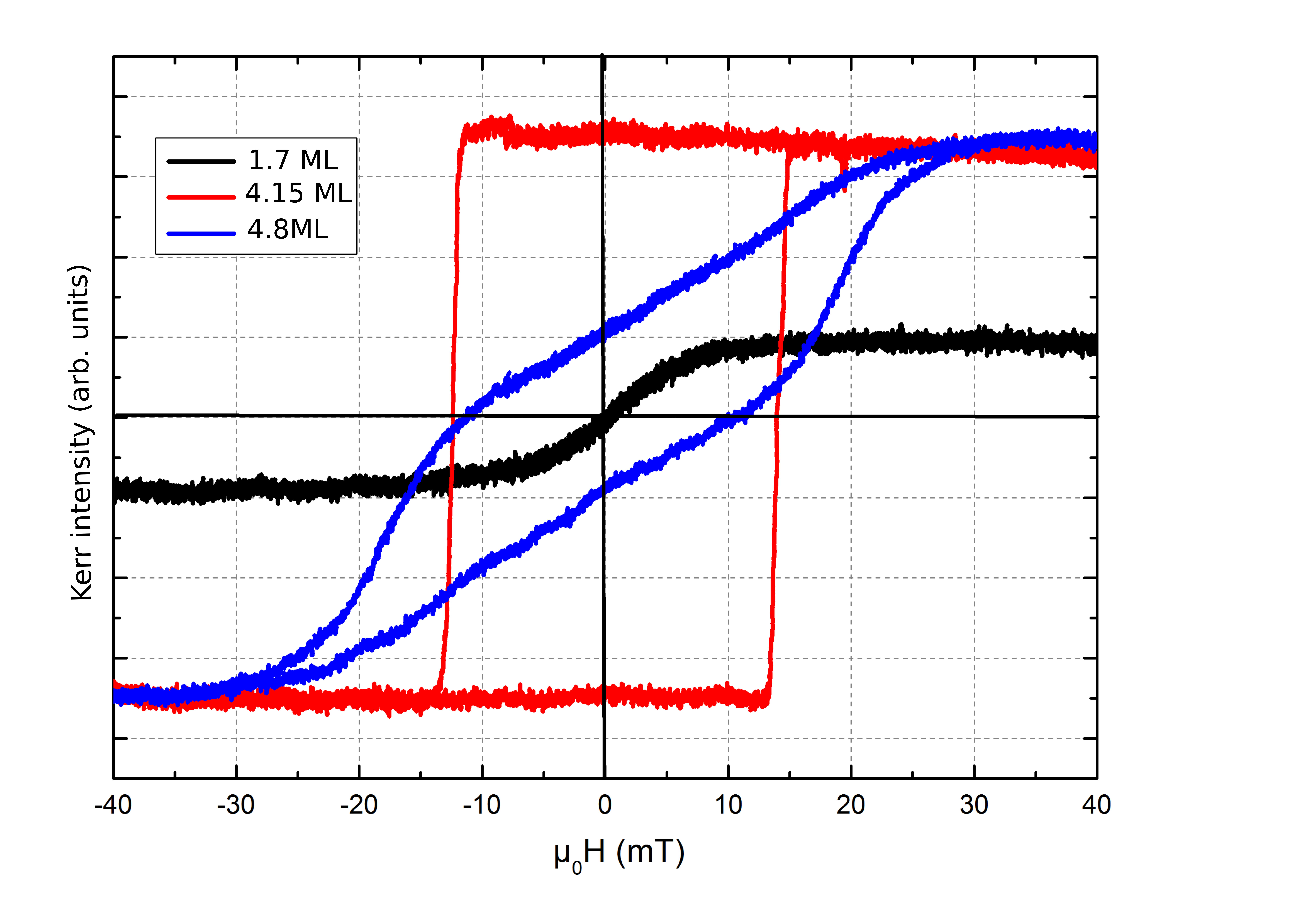}
  \end{center}
  \caption{\label{fig:kerr} Hysteresis loops obtained by polar focused Kerr on different positions along a Co wedge in Au/Co/W(110).}
\end{figure}

We used focussed Kerr magnetometry to study the magnetization reorientation in the sample with a Co wedge. The polar Kerr cycles are measured as a function of an out-of plane magnetic field. The hysteresis loops obtained for different Co thickness are plotted in Fig.~\ref{fig:kerr}. For small thickness the Curie temperature is below room temperature, i.e. the correlation between the Co atoms is weaker than the thermal fluctuations. In this regime the Co is paramagnetic and there is no Kerr signal \cite{Garreau1997}. The black loop in Fig.~\ref{fig:kerr} (1.7 ML) shows the presence of a finite magnetization, a saturation field of $\mu_0H_{\mathrm{sat}}= 10$ mT and the absence of coercivity. This indicates a superparamagnetic state where the Co islands are ferromagnetic with a weak mutual interaction. The out-of-plane interfacial Magneto-Crystalline Anisotropy is the dominating effect for the thickness range between $1.7$ ML and $4.5$ ML. Indeed the square hysteresis loop (red line in Fig.~\ref{fig:kerr}) for out-of-plane magnetic field shows the presence of an out-of-plane easy axis. The change of shape and the increase of saturation field in the hysteresis when increasing the thickness to $4.8$~ML (blue line in Fig.~\ref{fig:kerr}) show that the easy axis is not parallel to the applied field any longer. Indeed, when the thickness increases [$4.5-5.1$ ML] the magnetic volume increases and the shape anisotropy progressively tilts the magnetization in the sample plane. The spin reorientation range depends on the strength of the MCA and hence on the surface quality. The spin reorientation transition range is perfectly comparable with the values found by \cite{Sellmann2001b}. The complexity of the reflection mechanism in a multi-layer magnetic system does not allow to estimate from the Kerr magnetometry the value of the spontaneous magnetization ($M_s$). The spontaneous  magnetization ($M_s= 1.15\times10^6$ A/m) is obtained from the out-of-plane hysteresis loop taken using a Vibrating Sample Magnetometer and the estimation of the magnetic volume.

\section{Phenomenological interpretation of Anisotropic DMI}
The presence of anisotropic DMI in magnetic systems with $C_{2v}$ symmetry can phenomenologically be understood considering the Fert-Levy three atoms model \citep{FertLevy1980b}. This model considers a magnetic metal crystal and analyzes the indirect magnetic exchange between two magnetic atoms when the conduction electrons scatter with a high spin orbit coupling impurity. The indirect exchange interaction between two magnetic atoms via one electron in the conduction band is described by the RKKY model \citep{Ruderman1954}. In systems with inversion symmetry this interaction is symmetric, promotes collinear spin arrangement and its effect is hidden in the direct Heisenberg exchange. If we consider a system with lack of inversion symmetry and spin orbit coupling the symmetry nature of the indirect exchange strongly changes. Indeed the spin orbit coupling sets a relation between the space and spin degrees of freedom and in systems with lack of inversion symmetry a breaking of the exchange symmetry can thus be expected. In the Fert-Levy model the inversion symmetry is broken by the presence of a scattering point not collinear with the magnetic ions. The spin orbit coupling plays its role in the scattering process between the conduction electron and the impurity. The resulting exchange interaction thus will have a symmetric and an antisymmetric component. The antisymmetric one is the DMI-like interaction which promotes a perpendicular spin arrangement. Its Hamiltonian can be formulated \cite{FertLevy1980b} :

\begin{eqnarray}
\label{eq:DMI(r)}
H_{DM} &=& - V_1 \frac {
\sin [ \mathbf{k}_f( \mathbf{R}_1 + \mathbf{R}_2 + \mathbf{ R}_{12}) +( \pi / 10)Z] \hat{\mathbf{R}}_1 \cdot \hat{\mathbf{R}}_2 }{ \mathbf{R}_1 \mathbf{R}_2 \mathbf{R}_{12}}
( \hat{\mathbf{R}}_1 \times \hat{\mathbf{R}}_2) \cdot (\mathbf{S}_1 \times \mathbf{S}_2) \nonumber\\&=& \mathbf{d}(\mathbf{R}_1, \mathbf{R}_2, \mathbf{R}_{12}) \cdot (\mathbf{S}_1 \times \mathbf{S}_2)
\end{eqnarray}

where $V_1$ is the perturbation potential for the conduction electron gas that depends on the exchange matrix elements between a conduction electron and the d-orbital electrons in the magnetic atoms; in a framework set on the scattering center, $\mathbf{R}_1$ and $\mathbf{R}_2$ are the positions of the two magnetic atoms and $\mathbf{R}_{12}$ is the vector between the magnetic ions; the term $ (\pi / 10)Z$ is the Fermi level phase shift induced by the interaction of the conduction electron with the $Z$ electrons in the $d$ orbital of the scattering point and $\mathbf{k}_f$ is the wavevector of the conduction electron.\\
The strength and the sign of the interaction strongly depend on the geometry of the triangle composed by the ions and the scattering point. Indeed the DMI vector ($\mathbf{d}(\mathbf{R}_1, \mathbf{R}_2, \mathbf{R}_{12})$) has its direction always parallel to the normal of the triangle and its sign depends on the triangle geometry. In order to understand this dependence we analyze two different cases.
All the considerations will be extended to a two dimensional crystal where the $\mathbf{d}$ out-of-plane components are averaged to zero. Hence all the analysis will be developed in a 1D approximation.

\begin{itemize}
\item We study the DMI sign and strength fixing the distance between the line connecting the magnetic atoms and the scattering point, which is kept centered in between the magnetic atoms (Fig.~\ref{fig:DMI(r)}). The DMI strength in Fig.~\ref{fig:DMI(r)} is thus plotted as a function of the atomic distance R$_{12}$.

\begin{figure}[h]
 \begin{center}
 \includegraphics[width=8cm]{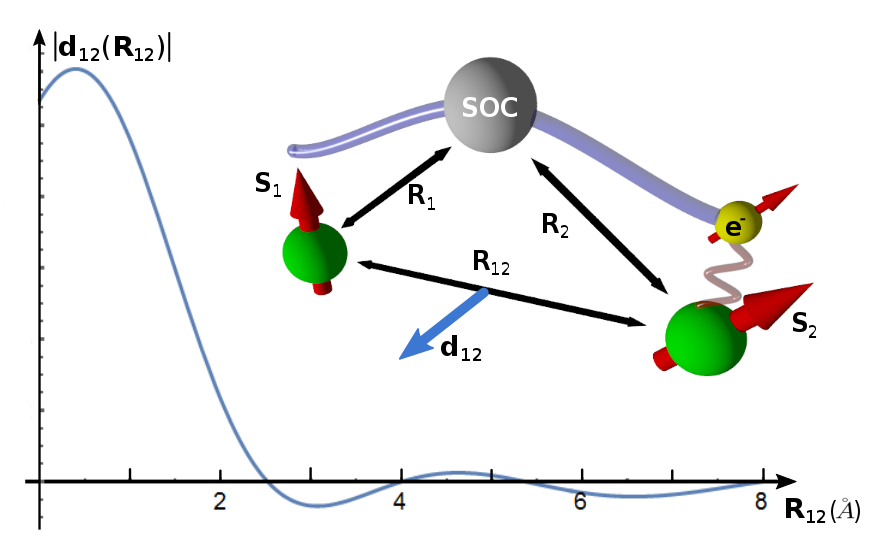}
   \end{center}
  \caption{\label{fig:DMI(r)} Plot of Eq.~(\ref{eq:DMI(r)}) as a function of the distance between the magnetic atoms $ \mathbf{R}_{12}$ in $\AA$. The y axis is normalized with respect to the constant $ V_1$. An artistic picture shows the configuration and the particles that play a role in the three atoms model for the DMI. }
\end{figure}

\item We fix the position of the magnetic atoms and change the scattering point position in a line parallel to the line connecting the magnetic atoms. The DMI strength in Fig.~\ref{fig:DMI(W)} is thus plotted as a function of the distance r.

\begin{figure}[h]
 \begin{center}
 \includegraphics[width=11cm]{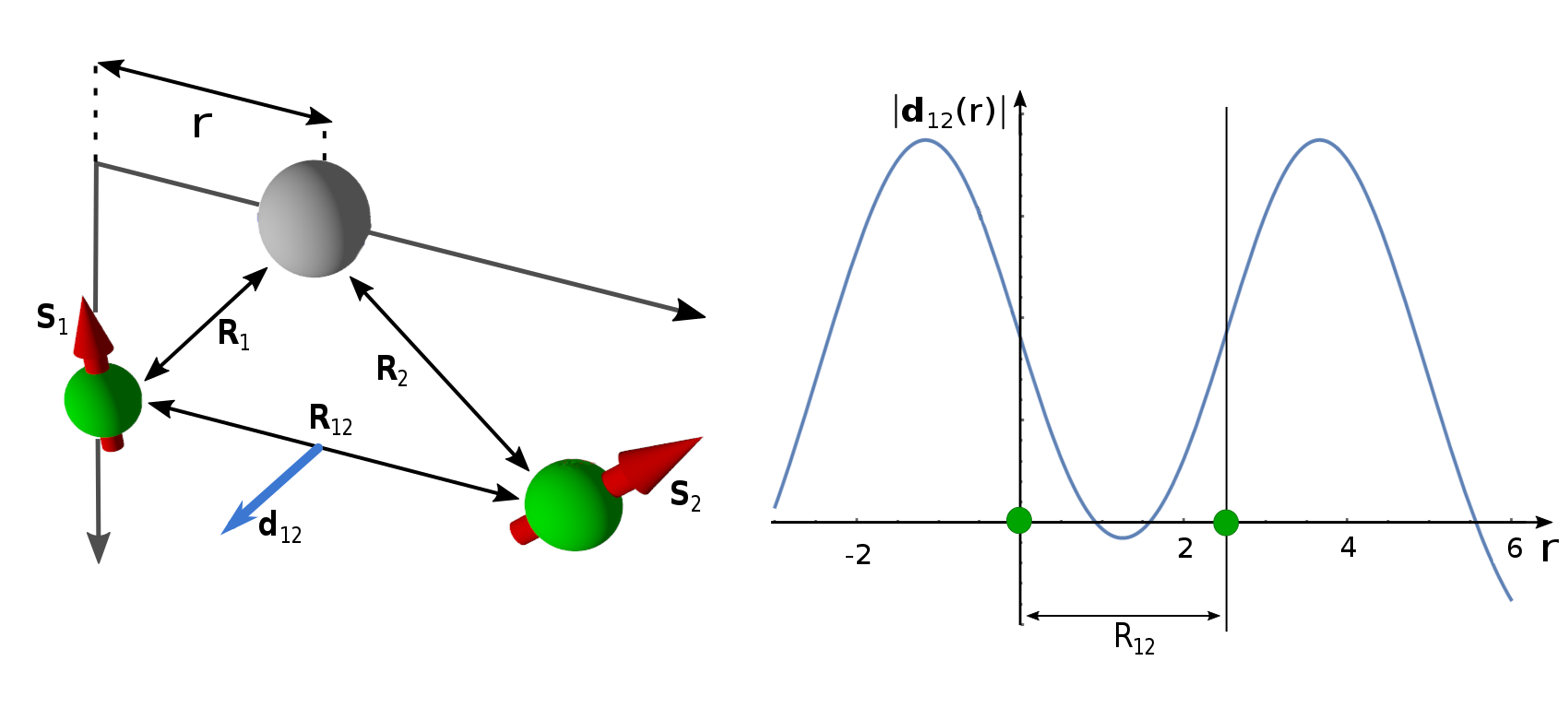}
   \end{center}
  \caption{\label{fig:DMI(W)} Plot of Eq.~(\ref{eq:DMI(r)}) as a function of the position of the scattering point $\mathbf{r}$. The y axis is normalized with  respect to the constant $ V_1$. The distance between the magnetic ions is set equal to $3 \AA$. }
\end{figure}

\end{itemize}

In both the analyzed cases $\mathbf{k}_f$ is estimated from angle-resolved photoemission spectroscopy measurements performed by Moras et al. \cite{Moras2015} on a Co/W(110) interface.
The plots in Fig.~\ref{fig:DMI(r)} and \ref{fig:DMI(W)} evidence the strong dependence of the DMI strength on the geometrical configuration. Indeed a change of the distance between the magnetic atoms or of the scattering point position drastically changes the interaction strength and can modify the DMI sign.\\

We consider the interface between a magnetic crystal and a heavy metal. In this stack the DMI arises from the interface with the high spin orbit coupling metal that breaks the inversion symmetry. The Fert-Levy model can not be used to have a quantitative interpretation of DMI strength and sign because the interaction can not be reduced to a simple scattering phenomenon. On the other hand we can use this model as a phenomenological tool for determining the crystal symmetry class where it is possible to expect an anisotropic DMI. Indeed we can consider the relationship between the two dimensional unit cells at the interface and take into account one by one the interactions between the magnetic ions with their closer scattering point independently one from the other.

We focus on the magnetic crystal and we consider the scattering points in the center of the bonds. In this approximation we can notice that a $C_{2v}$ crystal allows an anisotropic interaction. Indeed in a rectangular crystal the difference in distance between the two crystallographic directions could induce different strength and sign of the DMI vector. The same argument is not valid for higher symmetry classes like C$_{3v}$ or C$_{4v}$. In these crystals an anisotropic DMI can only be obtained if the scattering points are placed in a different way with respect to the bonds between magnetic atoms.

\section{From atomic structure to micromagnetic DMI in $C_{2v}$ symmetry systems}

The micromagnetic DMI is an averaged consequence of the atomic interactions. Hence in order to evidence the relationship between the crystal symmetry and the micromagnetic DMI it is fundamental to analyze the atomic configuration and symmetry of the interface between the magnetic and the heavy metal crystal. This analysis does not aim at the quantitative evaluation of DMI, but to illustrate how atomic DMI vectors $\mathbf{d}_{ij}$ between atoms i and j at various atomic sites add up to yield global micromagnetic DMI constants along the main symmetry axes of the system.
We consider the Co/W superstructure shown in Fig.~\ref{fig:super}.
The mismatch between the lattice parameters produces the relaxation of the Co crystal along the bcc[001] direction. In order to consider the full interface symmetry it is important to take into account the produced supercrystal described in Fig.~\ref{fig:super}, i.e the Co between 2ML and 10ML grows pseudomorphically along the $bcc[\overline{1}10]$ direction $(a_x = \sqrt{2}/2 a_W = 0.223$ nm) whereas along the  $bcc[001]$ it grows with a defined ratio with respect to the W ($a_y = 3.56/4.56 a_W/2  = 0.124$ nm) Fig.~\ref{fig:lattice} \cite{Fritzsche1995b}. It is possible to define a reconstruction period of $14 a_x$ where the Co crystal finds the initial relationship with the W ($14 a_x- 11 a_W=2 pm$) (Fig.~\ref{fig:super}).
In the Co/W superlattice the position of the W atoms with respect to the Co atoms changes from one Co unit cell to the next. Thus we can expect $\mathbf{d}_{ij}$ vectors with different strengths and directions.

\begin{figure}[h]
  \begin{center}
    \includegraphics[width=9cm]{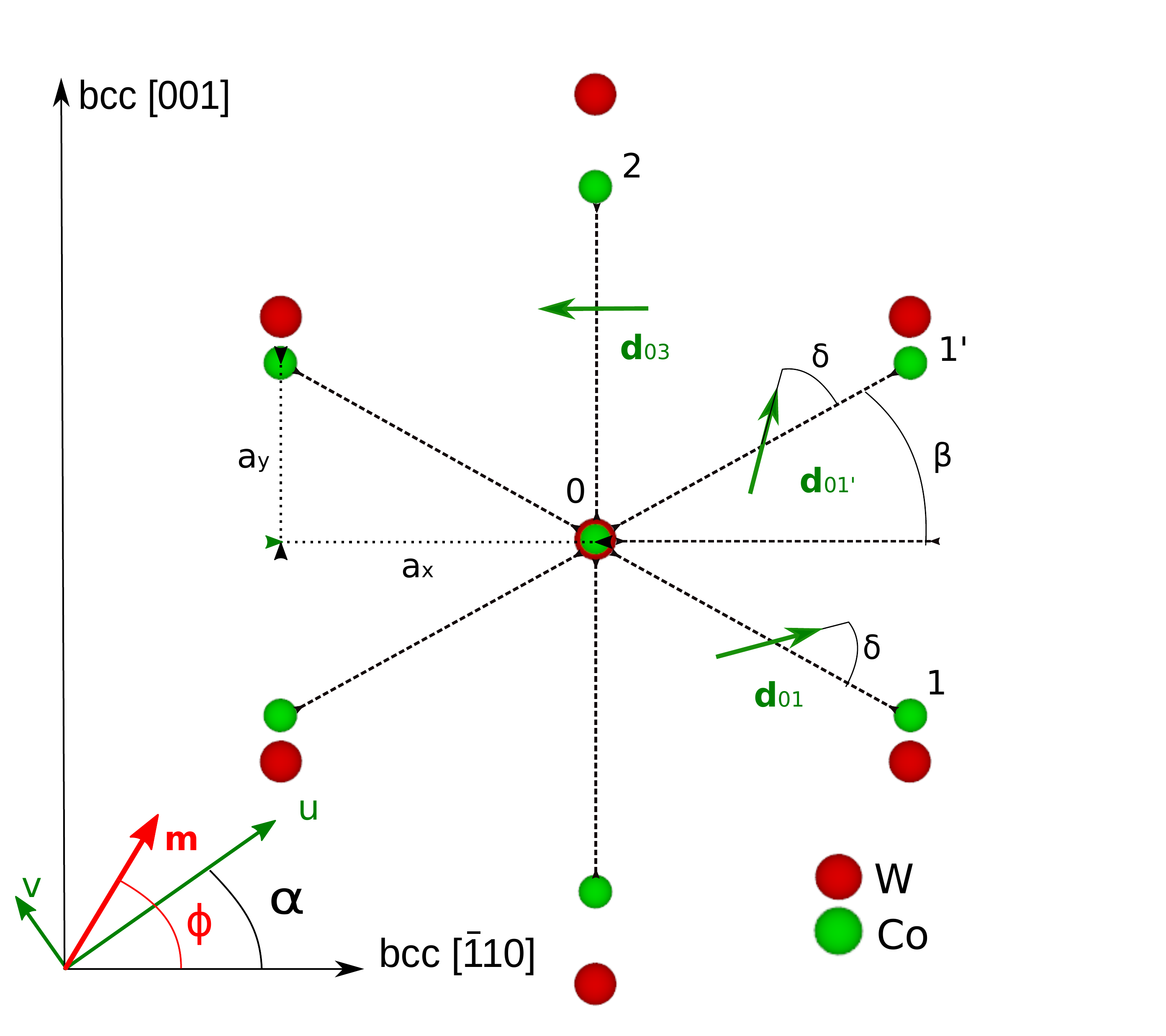}
  \end{center}
  \caption{\label{fig:lattice} Superposition of the W(110) and the strained Co (0001) surfaces with the Nishiyama-Wassermann relationship}
\end{figure}

If the analyzed magnetic configurations have a characteristic length ($l$) much larger than the supercell parameter ($14 a_x$) one can reduce the calculation to a single Co cell with  $\langle \mathbf{d}_{ij} \rangle$ vectors. These are  the average of all the $\mathbf{d}_{ij}$ for the same bonds on the superlattice.
In Fig.~\ref{fig:lattice} the higher symmetry Co/W cell with the $\langle \mathbf{d}_{ij} \rangle$ of the first neighbors are shown.

The vectors that describe the Co atoms positions with respect to the center Co $ \mathbf{a}_{ij}$ can be written in the crystal framework  $(\hat{x}// bcc[\overline{1}10], \hat{y}//bcc[001]) $:

\begin{eqnarray}
\mathbf{a}_{02}=& (0 , 2 a \sin \beta )\\
\mathbf{a}_{01^{\prime}}=& (a \cos \beta, - a \sin \beta ) \\
\mathbf{a}_{01}=& (a \cos \beta, a \sin \beta )
\end{eqnarray}

where $a$ is the length of the bond $01'$.
The Moriya symmetry rules \cite{Moriya1960b} allow to impose some constraints on the atomic $\langle \mathbf{d}_{ij} \rangle$ vectors.
Indeed along the $02$ bond the W atoms are aligned with the Co atoms and they belong to one of the mirror planes of the system.
Then the $\langle \mathbf{d}_{02} \rangle$ will lie in the crystal plane perpendicular to its bond.
Concerning  the bonds 01$^{\prime}$ and 01, it is possible to define a two-fold symmetry axis passing through the W atom position and perpendicular to these bonds.
Hence $\langle \mathbf{d}_{01^{\prime}} \rangle$ and $\langle \mathbf{d}_{01} \rangle$ will lie in the plane perpendicular to the two-fold axis, i.e. within the crystal plane.
The direction of the these DMI vectors in the plane is defined by two angles $\delta_{01^{\prime}}$ and $\delta_{01}$  Fig.~\ref{fig:lattice}.
It is possible to write these vectors in the two-fold crystal framework:

    \begin{eqnarray}
\langle \mathbf{d}_{01^{\prime}} \rangle &=& \left[ \langle d_{01^{\prime}} \rangle  \cos(\delta_{01^{\prime}} - \beta) ,
                                                    \langle d_{01^{\prime}} \rangle  \sin(\delta_{01^{\prime}} - \beta) +  ,
                                                   0 ) \right]\\
\langle \mathbf{d}_{01}          \rangle &=& \left[ \langle d_{01}          \rangle  \cos(\delta_{01}          + \beta) ,
                                                    \langle d_{01}          \rangle  \sin(\delta_{01}          + \beta),
                                                   0 \right]\\
\langle \mathbf{d}_{02}          \rangle &=& \left[ -\langle d_{y} \rangle , 0 , 0 \right]
 \end{eqnarray}

In a general two dimensional system the DMI energy  in the micromagnetic approach reads:

\begin{equation}
E_{DM} =  -  \int \left(  D_{xz,\mathrm{s}}^{(x)} L_{xz}^{(x)}+ D_{yz,\mathrm{s}}^{(y)} L_{yz}^{(y)} +D_{yz,\mathrm{s}}^{(x)} L_{yz}^{(x)} + D_{xz,\mathrm{s}}^{(y)} L_{xz}^{(y)}\right) dx dy
\end{equation}

where $D^{(i)}_{jk,\mathrm{s}}$ are the surfacic DMI micromagnetic constants (in J/m) and $L^{(i)}_{jk}$ are Lifshitz invariants $L^{(i)}_{jk}= m_j \frac{\partial m_k}{\partial i}-m_k \frac{\partial m_j}{\partial i} $ (where the indices $i$, $j$ and $k$ are the cartesian coordinates and $m^{(i)}$ the component of the magnetization along direction $i$).
The existence of this term $D^{(i)}_{jk,\mathrm{s}} L^{(i)}_{jk}$ is the signature that the DMI stabilizes a spin modulation where the j and k magnetic components change along the i component.
Along each direction the DMI stabilizes Bloch helicoids ( $D_{xz,\mathrm{s}}^{(y)} L_{xz}^{(y)}$ and $ D_{yz,\mathrm{s}}^{(x)} L_{yz}^{(x)} $) and N\'eel cycloids
($D_{xz,\mathrm{s}}^{(x)} L_{xz}^{(x)}$ and $ D_{yz,\mathrm{s}}^{(y)} L_{yz}^{(y)} $).
In a C$_{2v}$ symmetry system there are two mirror planes and the symmetry imposes that the magnetization rotates in these planes \citep{Moriya1960b}.
Hence along the main axes of the system the DMI stabilizes only N\'eel cycloids and its energy can be formulated in the main axis coordinate system :

\begin{equation}
\label{eq:EDMC2v}
E_{DM} =  -  \int \left( D_{xz,\mathrm{s}}^{(x)} L_{xz}^{(x)}+ D_{yz,\mathrm{s}}^{(y)} L_{yz}^{(y)}\right) dx dy
\end{equation}

Eq.~(\ref{eq:EDMC2v}) shows that the DMI micromagnetic constants stabilizing Bloch helicoids have to be zero.
This evidence can be used to set new constraints on the atomic $\mathbf{d}_{ij}$ vectors.

In order to understand how to set these constraints we develop a general model to elucidate how to pass from the atomic $\mathbf{d}_{ij}$ vectors to the micromagnetic $D^{(i)}_{jk,\mathrm{s}}$ constant for a system with different neighbors labeled with index k. Each bond can be characterized by the position of the Co atom $\mathbf{a}_k$ and by a DMI vector $\mathbf{d}_k $.
The atomic DMI energy in the first neighbors limit can be written:

\begin{equation}
\label{eq:Edmk}
E_{DM} = \frac12\sum_i\sum_{k\in \mathrm{NN}(i)} \mathbf{d}_k (\mathbf{m}_i \times \mathbf{m}_k)
\end{equation}

where the summations are respectively performed on all atoms $i$ and the nearest neighbors (NN) $k$ of $i$.
The case of a large length magnetic configuration allows to describe the magnetization in a continuous medium approach and to express $\mathbf{m}_k $ as the Taylor expansion of $\mathbf{m}$

\begin{eqnarray}
\label{eq:tayl}
\mathbf{m}(\mathbf{r}_k) &=&
        \mathbf{m}(\mathbf{r}_i)
        + (\mathbf{r}_k-\mathbf{r}_i)\cdot \mathbf{\hat{x}} \frac{\partial \mathbf{m}(\mathbf{r}_i)}{\partial x}
        + (\mathbf{r}_k-\mathbf{r}_i) \cdot \mathbf{\hat{y}} \frac{\partial \mathbf{m}(\mathbf{r}_i)}{\partial y}\nonumber\\
        &=&
        \mathbf{m}(\mathbf{r}_i)
        + a^{(x)}_k \frac{\partial \mathbf{m}(\mathbf{r}_i)}{\partial x}
        + a^{(y)}_k \frac{\partial \mathbf{m}(\mathbf{r}_i)}{\partial y}
\end{eqnarray}

then we can replace the Eq.~(\ref{eq:tayl}) in Eq.~(\ref{eq:Edmk}), developing the vectorial and scalar products and using the formalism of the Lifshitz invariants the DM energy can be written

\begin{equation}
\label{eq:4L}
E_{DM} = \frac12 \sum_{ik}\left[
  a^{(x)}_k d^{(x)}_k L^{(x)}_{yz}(\mathbf{r}_i)
- a^{(x)}_k d^{(y)}_k L^{(x)}_{xz}(\mathbf{r}_i)
+ a^{(y)}_k d^{(x)}_k L^{(y)}_{yz}(\mathbf{r}_i)
- a^{(y)}_k d^{(y)}_k L^{(y)}_{xz}(\mathbf{r}_i)
\right]
\end{equation}

Eq.~(\ref{eq:4L}) allows to calculate the micromagnetic DMI constant. Indeed we can transform the finite sum on the space into an integral on the unit cell.

\begin{equation}
\label{eq:4L_alt}
E_{DM} = \int d^2r\left[
  D^{(x)}_{yz,\mathrm{s}} L^{(x)}_{yz}(\mathbf{r}_i)
+ D^{(x)}_{xz,\mathrm{s}} L^{(x)}_{xz}(\mathbf{r}_i)
+ D^{(y)}_{yz,\mathrm{s}} L^{(y)}_{yz}(\mathbf{r}_i)
+ D^{(y)}_{xz,\mathrm{s}} L^{(y)}_{xz}(\mathbf{r}_i)
\right]
\end{equation}
with the micromagnetic DMI constants

\begin{subequations}
    \begin{eqnarray}
\label{eq:Dmicro}
D^{(x)}_{xz,\mathrm{s}} =  \frac{1}{2S} \sum_k a^{(x)}_k d^{(y)}_k
\qquad
D^{(y)}_{yz,\mathrm{s}} = -\frac{1}{2S} \sum_k a^{(y)}_k d^{(x)}_k
\\
\label{eq:cond}
D^{(y)}_{xz,\mathrm{s}} =  \frac{1}{2S} \sum_k a^{(y)}_k d^{(y)}_k
\qquad
D^{(x)}_{yz,\mathrm{s}} = -\frac{1}{2S} \sum_k a^{(x)}_k d^{(x)}_k
    \end{eqnarray}
\end{subequations}
where $S$ is the unit cell surface. Then if the analyzed system has a C$_{2v}$ symmetry $D^{(y)}_{xz,\mathrm{s}} = D^{(x)}_{yz,\mathrm{s}} = 0 $ and these relations can be used to set the relationship between the $\mathbf{d}_{ij}$ vectors. For the sake of simplicity in the main manuscript the $D^{(x)}_{xz,\mathrm{s}} $ and $D^{(y)}_{yz,\mathrm{s}} $ constants are renamed respectively $D^{(x)}_{\mathrm{s}} $ and $D^{(y)}_{\mathrm{s}} $.\\

Now we can find the micromagnetic $D_s$ constants and their relationship with the atomic $\mathbf{d}_{ij}$ for the case of Co/W(110).
Replacing the $\mathbf{a}_k $ and the $\mathbf{d}_k$ in the Eq.~(\ref{eq:cond}) and setting them equal to zero we obtain:

\begin{equation}
a \cos \beta \langle d_{01} \rangle \cos( - \beta + \delta_{01}) + a \cos \beta  \langle d_{01^{\prime}} \rangle \cos(  \beta + \delta_{01^{\prime}}) = 0
\end{equation}
\begin{equation}
- a \sin \beta \langle d_{01} \rangle \sin( - \beta + \delta_{01}) + a \sin \beta  \langle d_{01^{\prime}} \rangle \sin(  \beta + \delta_{01^{\prime}}) = 0
\end{equation}

solving the system we find that $\langle d_{01} \rangle=  \langle d_{01^{\prime}} \rangle = d $ and $\delta_{01}= \pi - \delta_{01^{\prime}} $
Hence the micromagnetic $D_s$ constants become:

\begin{equation}
D^{(x)}_{\mathrm{s}} = \frac{d}{a} \frac{\sin( \beta + \delta_{01})}{\sin \beta}
\end{equation}

\begin{equation}
D^{(y)}_{\mathrm{s}} = \frac{2d}{a} \left[ \frac{\cos( \beta + \delta_{01})}{\cos \beta}  -\frac{d_y}{d} \frac{1}{\cos \beta} \right]
\end{equation}

\begin{figure}[h]
  \begin{center}
    \includegraphics[width=9cm]{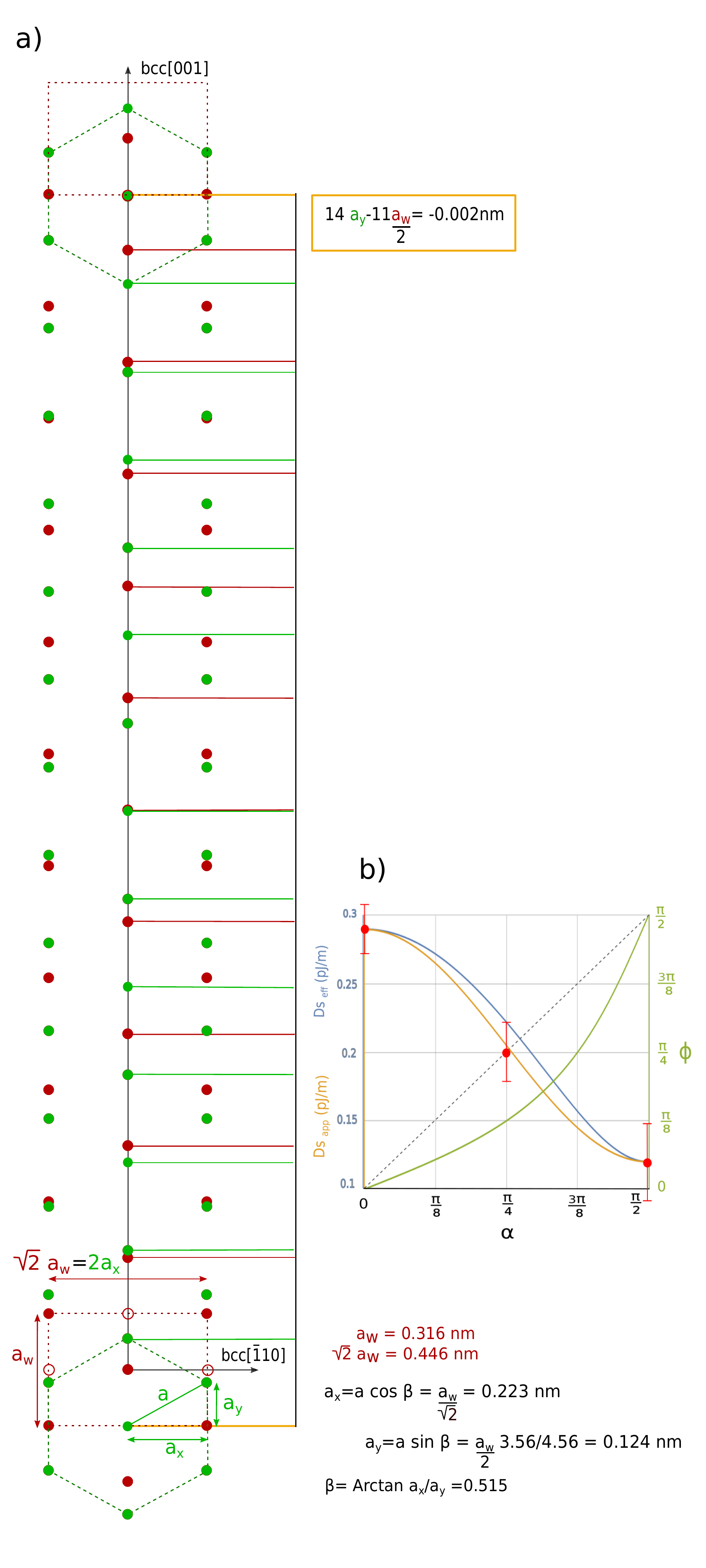}
  \end{center}
  \caption{\label{fig:super}\textbf{(a)} Sketch of the supercrystal produced by the epitaxial growth of the Co(0001) surface on the bcc W(110) surface. \textbf{(b)} Blue and orange lines : micromagnetic calculated $D^{(\mathrm{eff})}_{\mathrm{s}}$(Eq.~(\ref{eq:DeffS})) and $D^{(\mathrm{app})}_{\mathrm{s}}$(Eq.~(\ref{eq:DappS}))  as a function of the in-plane directions ($\phi$); red dots: D strength evaluated from the experimental data; green line: micromagnetic calculated magnetization promoted by DMI (Eq.~(\ref{eq:phialpS})) as a function of the crystallography directions; dashed line: N\'eel-like cycloid  }
\end{figure}

\section{Spin texture and effective DMI in a $C_{2v}$ symmetry system}

It is possible to generalize the discussion above to every ultra-thin magnetic film with interfacial DMI and a twofold symmetry. Considering a new basis $(\widehat{u} ,\widehat{v},\widehat{z})$, turned by an angle $\alpha = (\widehat{x},\widehat{u})$ around the vertical axis (Fig.~\ref{fig:lattice}), with respect to the initial basis, the DMI energy of a one-dimensional spin modulation propagating along $\widehat{u}$ reads:
\begin{align}
E_{DM}(\alpha) =
 -  \int   \left[ \cos^{2}(\alpha) D^{(x)}_{\mathrm{s}} + \sin^{2}(\alpha) D^{(y)}_{\mathrm{s}} \right]  L_{uz}^{(u)}  d^2r
 -  \int   \left(  D^{(x)}_{\mathrm{s}}- D^{(y)}_{\mathrm{s}}  \right) \cos(\alpha)\sin(\alpha)          L_{vz}^{(u)}  d^2r.
\end{align}
It presents two different types of Lifshitz invariants that describe a DMI stabilizing different spin configurations in competition. It means that in a general two-fold system the DMI promotes N\'eel cycloids along the main axes and a mixed configuration between a N\'eel cycloid and a Bloch helicoid along the intermediate directions.

In order to understand the BLS data at $\alpha=\pi/4 $, we evaluate the DMI energy for a SW described as $\mathbf{m}(u,t)= \mathbf{M} + \delta \mathbf{m}(u,t)$, with $\mathbf{M}=\sqrt{1-\delta m^2}\,\widehat{v}$ (parallel to $\mathbf{H}_{\mathrm{ext}}$, due to the DE geometry) and $\delta \mathbf{m}(u,t)=\delta m[\sin(k_{sw}u-\omega t)\widehat{u}+\cos(k_{sw}u-\omega t)\widehat{z}]$ lying in the $(\widehat{u},\widehat{z})$ plane, with $k_{sw}$ the wave vector magnitude and $\delta m$ the spin wave amplitude. The DMI energy density $\omega_{DM}^{sw}$ of the texture is
\begin{equation}
    \omega_{DM}^{sw}=\frac{2\pi}{\Lambda}\left[ \cos^{2}(\alpha) D^{(x)}_{\mathrm{s}} + \sin^{2}(\alpha) D^{(y)}_{\mathrm{s}} \right]\delta m^2.
\end{equation}
with $\Lambda$ the spinwave wave length. The $2\pi/\Lambda$ factor arises from the fact that over one period, the varying part of the magnetization undergoes a $2\pi$ rotation. As the spin-wave has a N\'eel structure, only the energy part associated to $L_{uz}^{(u)}$ remains. Therefore, an apparent DMI constant $D^{(app)}_{\mathrm{s}}$ is estimated as $D^{(app)}_{\mathrm{s}}=\omega_{DM}^{sw}\Lambda/2\pi\delta m^2$. As a function of $\alpha$, we find
\begin{equation}
 D^{(app)}_{\mathrm{s}}= D^{(x)}_{\mathrm{s}} \cos^2 \alpha + D^{(y)}_{\mathrm{s}} \sin^2 \alpha   \label{eq:DappS}
\end{equation}
The plot of Eq.~(\ref{eq:DappS}) in Fig.~\ref{fig:super}(b) shows a very good agreement with the experimental data.

In a spin spiral along $\widehat{u}$, the modulation plane is free to rotate around the vertical axis in order to further minimize the energy. Writing $\phi$ the angle of such a plane with respect to the $\widehat{x}$ axis (Fig.~\ref{fig:lattice}), the spin spiral is expressed in the $(\widehat{x},\widehat{y},\widehat{z})$ plane as $\mathbf{m}(u)=\{\sin(2\pi u/\Lambda')\cos\phi;\sin(2\pi u/\Lambda')\sin\phi;\cos(2\pi u/\Lambda')\}$ with $\Lambda'$ the spiral wave length. Minimizing the energy (Eq.~(\ref{eq:EDMC2v})) with respect to $\phi$, we calculate the spin spiral plane orientation as a function of $\alpha$:
\begin{align}
 \tan\phi =& \left( \frac{ D^{(y)}_{\mathrm{s}}}{ D^{(x)}_{\mathrm{s}}} \right) \tan\alpha \label{eq:phialpS}
 \end{align}
This result shows that along the main axis ($\alpha = 0$ or $\pi/2$), $\phi = \alpha$ and N\'eel cycloids are stabilized, whereas along arbitrary directions mixed solutions are found, except if $D^{(y)}_{\mathrm{s}}= D^{(x)}_{\mathrm{s}}$. In the particular case of $D^{(y)}_{\mathrm{s}}= -D^{(x)}_{\mathrm{s}}$, $\phi = -\alpha$ and pure Bloch helicoids are found when $\alpha = \pi/4 + 2n/\pi$.
From the DMI energy density of this spin spiral $\omega_{DM}^{ss}$, it is possible to define an effective DMI constant as  $D^{(eff)}_{\mathrm{s}}=\omega_{DM}^{ss}\Lambda/2\pi$

\small
\begin{align}
D_s^{eff}=   D^{(x)}_{\mathrm{s}} \cos\alpha  \cos\left[ \arctan\left( \frac{ D^{(y)}_{\mathrm{s}}}{ D^{(x)}_{\mathrm{s}}} \tan \alpha \right)\right]
+  D^{(y)}_{\mathrm{s}} \sin\alpha \sin\left[ \arctan \left( \frac{ D^{(y)}_{\mathrm{s}}}{ D^{(x)}_{\mathrm{s}}}\tan \alpha \right )\right] \label{eq:DeffS}
 \end{align}

\end{widetext}

\end{document}